\documentclass{emulateapjx}

\usepackage{amsmath}

\newcommand{\xc}{{x^{(2)}}}
\newcommand{\detg}{{\sqrt{-g}}}
\newcommand{\dof}{{\rm dof}}
\newcommand{\LP}{{\rm LP}}
\newcommand{\CP}{{\rm CP}}
\newcommand{\EVPA}{{\rm EVPA}}

\begin{document}
\shorttitle{Sgr A* Modeling by 3D GRMHD and Polarized Transfer} \shortauthors{Shcherbakov, Penna \& McKinney}

\title{SAGITTARIUS A* ACCRETION FLOW AND BLACK HOLE PARAMETERS \\FROM GENERAL RELATIVISTIC DYNAMICAL AND POLARIZED RADIATIVE MODELING}
\author{Roman V. Shcherbakov\altaffilmark{1}, Robert F. Penna\altaffilmark{2}, Jonathan C. McKinney\altaffilmark{3}}
\altaffiltext{1}{\url{http://astroman.org} \linebreak Department of Astronomy, University of Maryland, College Park, MD 20742-2421, USA; Hubble Fellow}
\altaffiltext{2}{rpenna@cfa.harvard.edu \linebreak Harvard-Smithsonian Center for Astrophysics, 60 Garden Street, Cambridge, MA 02138, USA}
    \altaffiltext{3}{jmckinne@stanford.edu \linebreak Department of Physics and Kavli Institute for Particle Astrophysics and Cosmology, Stanford University, Stanford, CA 94305-4060, USA; Chandra Fellow}
\email{roman@astro.umd.edu}
\begin{abstract}

  We obtain estimates of Sgr A* accretion flow and black hole parameters by fitting polarized sub-mm observations with
  spectra computed using three-dimensional (3D) general relativistic (GR) magnetohydrodynamical (MHD) (GRMHD) simulations.
  Observations are compiled from averages over many epochs from
  reports in $29$ papers for estimating the mean fluxes $F_\nu$,
  linear polarization (LP) fractions, circular polarization (CP)
  fractions, and electric vector position angles (EVPAs).  GRMHD simulations are computed with dimensionless spins
  $a_*=0,0.5,0.7,0.9,0.98$ over a $20,000M$ time interval.  We perform
  fully self-consistent GR polarized radiative transfer using our new code to explore the
  effects of spin $a_*$, inclination angle $\theta$, position angle (PA),
  accretion rate $\dot{M}$, and electron temperature $T_e$ ($T_e$ is reported for radius
  $6M$). By fitting the mean sub-mm fluxes and LP/CP fractions, we obtain estimates for these model parameters
  and determine the physical effects that could produce polarization signatures.
  Our best bet model has $a_*=0.5$, $\theta=75^\circ$, ${\rm PA}=115^\circ$,
  $\dot{M}=4.6\times10^{-8}M_\odot{\rm year}^{-1}$, and $T_e=3.1\times10^{10}$~K at $6M$.
  The sub-mm CP is mainly produced by Faraday conversion as modified by Faraday rotation, and
  the emission region size at $230$~GHz is consistent with the VLBI size of $37\mu$as.
  Across all spins, model parameters are in the ranges $\theta=42^\circ-75^\circ$,
  $\dot{M}=(1.4-7.0)\times10^{-8}M_\odot{\rm year}^{-1}$, and  $T_e=(3-4)\times10^{10}$K.
  Polarization is found both to help differentiate models and to introduce new
  observational constraints on the effects of the magnetic field
  that might not be fit by accretion models so-far considered.

\end{abstract}
\keywords{accretion, accretion disks –-- black hole physics –-- Galaxy: center --– radiative transfer –-- relativistic
processes --- polarization}

\section{INTRODUCTION}

The mass of the Galactic Center black hole (BH) is $M\approx
4.5\cdot10^6 M_\odot$ \citep{ghez,reid08,gillessen} and the spin is uncertain
\citep{huangspin,broderick,moscibr_sim,broderick10,dexter2010}.  It
resides at a distance $d\approx 8.4$~kpc.  Because of its proximity,
it has been observed in many wavelengths: $\gamma$-rays, X-rays, IR,
(sub-)mm, and radio. X-ray bremsstrahlung emission originates from hot
gas at large radii where the BH's gravity becomes important
\citep{narayan_nat,narayan98,shcher_cond} and Compton-scattered
emission originates from close to the horizon \citep{moscibr_sim}. X-rays at large
radii are spatially resolved and have been used to constrain
dynamical models for this region \citep{shcher_cond}. The sub-mm
emission is cyclo-synchrotron emission originating from close to the
BH. Cyclo-synchrotron emission is polarized, and both linear and
circular polarizations have been observed from Sgr A* at several
sub-mm wavelengths. The accretion flow was recently resolved at
$230$~GHz \citep{doeleman,fish10}. General relativistic (GR) effects
were deemed necessary to explain the small size with full width at
half maximum (FWHM) of $37\mu$as.  Radio emission is also produced by
cyclo-synchrotron at larger distances from the BH.  Relativistic
frame-dragging is important near the BH, so sub-mm polarized
observations and the Compton-scattered X-rays might help to constrain
the BH spin.  The goal of the present paper is to model the sub-mm
in the range of $88$~GHz to $857$~GHz in order to estimate the accretion
flow and black hole parameters.

Sgr A* is a variable source with a variability amplitude routinely
reaching $30\%$ in sub-mm. A popular approach is to fit simultaneous
observations (e.g. \citealt{yuan_data,broderick}), in particular, the
set from \citet{falcke98}. However, in such an approach, one would use
a single simultaneous set of observations. However, simultaneous
observations of fluxes, linear polarization (LP), and circular
polarization (CP) fractions at several frequencies are not
available. So we consider non-simultaneous statistics of all
observations at all frequencies and find the mean values and standard
errors of quantities at each frequency.

Numerous accretion flow models have been applied to the Galactic
Center: advection-dominated accretion flow (ADAF) \citep{narayan},
advection-dominated inflow-outflow solution (ADIOS) \citep{blandford},
jet-ADAF \citep{yuan_jet}, jet \citep{maitra}, and viscous and
magnetohydrodynamical (MHD) numerical simulations. The
quasi-analytical models are useful because there is little expense in
changing parameters.  However, they have a large number of free
parameters and also incorporate many assumptions that are not
justifiable from first principles \citep{huang,huangnew}, which leads
to systematic uncertainties in all fits. Numerical simulations require
fewer inputs and are useful for more quantitative modeling of the
plasma near a rotating BH.  General relativistic (GR) MHD
(GRMHD) simulations (especially three-dimensional (3D) simulations),
which are run over a sufficiently long duration, are still computationally
expensive and involve state-of-the-art codes that are still being
developed \citep{mb09,fragile,noble09,moscibr_sim,penna}.  Yet,
these expensive 3D simulations are required to model the turbulent disk flow,
because 2D axisymmetric simulations cannot sustain turbulence
as shown by generalizations of Cowling's anti-dynamo theorem \citep{hide82}.
Given their expense, such 3D GRMHD simulations are limited to a region
relatively close to the BH \citep{dexter,moscibr_sim}, whereas some
emission and some Faraday rotation might happen far from the BH. So we
analytically extend the modeled region out to $20,000M$, perform
radiative transfer, and find the best fit to the data. The extension
to large radius allows us to define the electron temperature more
consistently \citep{sharma_heating}. We find a posteriori (see
Appendix~\ref{s_tests}) that the simulated polarized spectra are not
overly sensitive to the details of the analytic extensions of density
and temperature, but may depend on the extension of the magnetic
field.

The radiation close to the BH has been modeled in Newtonian
\citep{yuan_data} and quasi-Newtonian approximations
\citep{goldston,CK}. It has been modeled in GR assuming unpolarized
\citep{fuerst,dexter,dolence} and polarized
\citep{broderick,shcher_huang} light.  Fitting the total flux spectrum
might not be sufficient to estimate the spin, and naturally one
expects polarization to provide extra observational constraints. Spin values from
$a_*=0$ \citep{broderick} to $a_*=0.9$ \citep{moscibr_sim} have been
estimated. We neglect Comptonization \citep{moscibr_sim} and radiation
from non-thermal electrons \citep{mahadevan,ozel,yuan_data}.
Emissivities are calculated in the synchrotron approximation
\citep{legg,sazonov,pach,melrose_emis} with an exact thermal electron
distribution.  Discrepancies with the exact cyclo-synchrotron
emissivities \citep{leung,shcher_huang} are negligible as estimated in
\S~\ref{s_transf}.  Exact Faraday rotation and conversion expressions
are used \citep{shcher_farad}.

We compare simulated spectra to observed ones at many frequencies
simultaneously, extending an approach pioneered by \citet{broderick}
and \citet{dexter}.  We compute the average observed spectra, find the
deviations of the means, and then compare them to the average simulated
spectra.  In the search for the best fit models, we are guided by
the value of $\chi^2/\dof$, which is the normalized sum of squares of
normalized deviations.  Yet, we leave the exploration of the
statistical meaning of $\chi^2/\dof$ to future work.  We search the
space of all parameters: spin $a_*$, inclination $\theta$, ratio of
proton to electron temperatures $T_p/T_e$ ($T_p/T_e$ is reported for radius
$6M$), and accretion rate $\dot{M}$ to find the minimum $\chi^2$
models.

We summarize the radio/sub-mm observations of Sgr A* in \S~\ref{s_obs}.
Our 3D GRMHD simulations are described in
\S~\ref{s_simul} together with the physically-motivated extension to
large radii, and the electron heating prescription. We run simulations
for dimensionless spins $a_* = a/M = 0, 0.5, 0.7, 0.9, 0.98$. The GR
polarized radiative transfer technique is described in
\S~\ref{s_transf}.

The set of observations we consider consists of
the spectral energy distribution (SED) within the $88$~GHz to
$857$~GHz frequency range, linear polarization (LP) fractions at
$88$~GHz, $230$~GHz, and $349$~GHz, and circular polarization (CP)
fractions at $230$~GHz and $349$~GHz.  In \S~\ref{s_result} we discuss
our results: the best fit models to the observations,
the importance of various physical effects in producing the observed CP and LP
and electric vector position angle (EVPA), and image size estimates.
We produce the simulated images of total and polarized intensities.
Discussion in \S~\ref{s_discus} compares the
results to previous estimates, emphasizes the significance of
polarization, notes the sources of error, and outlines prospects for
future work. In Appendix~\ref{s_tests} we describe a number of
convergence tests of our GR polarized radiative transfer code and the
radial extension of the dynamical model. Throughout the paper we
measure distance and time in the units of BH mass $M$ by setting the
speed of light $c$ and gravitational constant $G$ to unity.

\section{OBSERVATIONS}\label{s_obs}

Sgr A* is known to be a highly variable source, yet quiescent models
of Sgr A* emission are popular and useful. Unlike the drastic
variations of X-ray and NIR fluxes \citep{baganoff01,genzel2003},
sub-mm fluxes do not vary by more than a factor of $2-3$
\citep{zhao2003}.  We compile the set of observed polarized fluxes
at each frequency, then we find the mean spectrum and the errors of the mean fluxes.

Previously, the observed flux spectra were compiled by
\citet{yuan_data,broderick}. However, both papers summarize a limited
set of observations and concentrate on simultaneously observed fluxes. Sub-mm flux data
reported in \citet{yuan_data} consist of a short set of observations
by \citet{falcke98} and one set of SMA observations by
\citet{zhao2003}. \citet{broderick} adds to these the rest of SMA
total flux data
\citep{marrone2006a,marrone2006b,marrone,marrone2008}. So $6$ out of
at least $29$ papers on sub-mm observations of Sgr A* were
taken into account. We compute an averaged spectrum based on $29$ papers
reporting sub-mm observations of Sgr A*.

\begin{table*}
\scriptsize
\caption{Summary of Sgr A* radio/sub-mm observations}\label{tab_obs}
\begin{tabular}{ | p{10mm} | p{20mm}| p{59mm} | p{20mm} | p{17mm} | p{18mm} | }
\tableline\tableline
  $\nu$ [GHz]  & Telescopes & $F_\nu$ [Jy] & LP [$\%$] & CP [$\%$] & EVPA [$^\circ$] \\
\tableline
  {\bf 8.45} & VLA &{$\bf 0.683\pm0.032$} \citep{serabyn,falcke98,bower99a,an} & \nodata & {$\bf -0.26\pm0.06$\tablenotemark{a}} \citep{bower99a}&  \\\tableline
  {\bf 14.90} & VLBA, VLA&{$\bf 0.871\pm0.012$} \citep{serabyn,falcke98,bower2002, herrnstein,an,yusef-zadeh} & \nodata & {$\bf -0.62\pm0.26$\tablenotemark{a}} \citep{bower2002}&  \nodata\\\tableline
  {\bf 22.50} & VLBA, VLA&{$\bf 0.979\pm0.016$} \citep{serabyn,falcke98,bower99b,herrnstein,an,lu2008,yusef2007,yusef-zadeh} & {$\bf 0.20\pm0.01$\tablenotemark{a}} \citep{bower99b,yusef2007} & \nodata &  \nodata\\\tableline
  {\bf 43} & GMVA, VLBA, VLA&{$\bf 1.135\pm0.026$} \citep{falcke98,lo98,bower99b,herrnstein,an,shen,krichbaum,yusef2007,lu2008,yusef-zadeh}& {$\bf 0.55\pm0.22$\tablenotemark{a}} \citep{bower99b,yusef2007} &  \nodata&\nodata  \\\tableline
  {\bf 88} & BIMA, MPIfR, VLBA, VLA, Nobeyama, NMA, CARMA&{$\bf 1.841\pm0.080$} \citep{falcke98,krich98,bower99b,doele2001,miyazaki,shen,krichbaum,macquart,lu2008,yusef-zadeh}& {$\bf 1.42\pm0.5$\tablenotemark{a,b}} \citep{bower99b,macquart}& \nodata & {\bf -4\tablenotemark{c}} \citep{bower99b,shen,macquart}\\\tableline
  {\bf 102} & OVRO, CSO-JCMT, Nobeyama, NMA, IRAM &{$\bf 1.91\pm0.15$} \citep{serabyn,falcke98,miyazaki,mauerhan,yusef-zadeh} & \nodata& \nodata & \nodata \\\tableline
  {\bf 145} & Nobeyama, NMA, IRAM, JCMT &{$\bf 2.28\pm0.26$} \citep{falcke98,aitken2000,miyazaki,yusef-zadeh} & \nodata & \nodata & \nodata \\\tableline
  {\bf 230} & IRAM, JCMT, BIMA, SMA, OVRO &{$\bf 2.64\pm0.14$} \citep{serabyn,falcke98,aitken2000,bower2003,bower2005,zhao2003,krichbaum,marrone2006a,
marrone,marrone2008,doeleman,yusef-zadeh}& {$\bf 7.40\pm0.66$} \citep{bower2003,bower2005,marrone,marrone2008}& {$\bf -1.2\pm0.3$\tablenotemark{a}} (\citet{munoz_poster,munoz}) & {$\bf 111.5\pm5.3$} \citep{bower2003,bower2005,marrone,marrone2008}\\\tableline
  {\bf 349} & SMA, CSO, JCMT &{$\bf 3.18\pm0.12$} \citep{aitken2000,an,marrone2006b,marrone,marrone2008,yusef-zadeh}& {$\bf 6.50\pm0.61$} \citep{marrone2006b,marrone} & {$\bf -1.5\pm0.3$ \tablenotemark{a}} (\citet{munoz}) & {$\bf 146.9\pm2.2$} \citep{marrone2006b,marrone} \\\tableline
  {\bf 674} & CSO, SMA &{$\bf 3.29\pm0.35$} \citep{marrone2006a,marrone2008,yusef-zadeh}& \nodata & \nodata & \nodata \\\tableline
  {\bf 857} & CSO &{$\bf 2.87\pm0.24$} \citep{serabyn,marrone2008,yusef-zadeh}& \nodata & \nodata & \nodata \\\tableline
\end{tabular}
\tablenotetext{1}{The uncertainty of the mean of these quantities is given by instrumental errors.}
\tablenotetext{2}{The mean LP at $3.5$~mm is computed based on lower and upper sidebands in \citet{macquart}. The error is based on $0.5\%$ systematic error reported therein.}
\tablenotetext{3}{The mean EVPA at $88$~GHz is uncertain due to $\pm180^\circ$ degeneracy; e.g. the reported $\EVPA=80^\circ$ could as well be interpreted as $-100^\circ$.}
\end{table*}

The reported observations vary in covered period from several hours \citep{an}
to several years \citep{zhao2003,krichbaum}.  We know that variations
of a factor of $2$ may happen within several hours
\citep{yusef-zadeh}, whereas variations by more than a factor of
several are never observed in the sub-mm.  So, fluxes observed more
than a day apart are weakly correlated. The issue of autocorrelation in
timescales will be addressed in future work. We consider the following
averaging technique to sample the distributions of fluxes. First, we
define groups of close frequencies, the frequencies in each group
being different by no more than several percent from the mean. There
are $11$ such groups (see Table~\ref{tab_obs}). We exclude papers
reporting single frequencies far from the mean of each group. In
particular, the $94$~GHz and $95$~GHz observations of
\citet{li2008,falcke98} and the $112$~GHz observations of
\citet{bower2001} are excluded. A mean frequency is ascribed to
represent each group. Then, we take all reported observations of each
polarization type (total flux, LP and CP fraction, EVPA) for each
group and draw the largest sample of fluxes/polarization fractions,
taking observations separated by at least $24$ hours. When several
fluxes are reported over a period of several hours
\citep{yusef-zadeh}, we draw one data point from the very
beginning of the observation, unless a flare is reported to occur at that time.
Some of the published observations have large error bars. Often such data
are produced by observing in sub-mm with large beam size,
but light from Sgr A* is blended with dust and other
sources. In particular, SMT data \citep{yusef-zadeh}, early CSO
measurements \citep{serabyn}, and early JCMT measurements
\citep{aitken2000} may have such issues, so we exclude these data from
the sample. The interferometric observations, especially with VLBI,
help to reduce the error from otherwise unreliable observations,
e.g. with BIMA array \citep{bower2001}. However, some inconsistencies
still exist for simultaneous observations at the same frequency with
different instruments \citep{yusef-zadeh}.

\begin{figure*}[ht]\epsscale{1}
\plotone{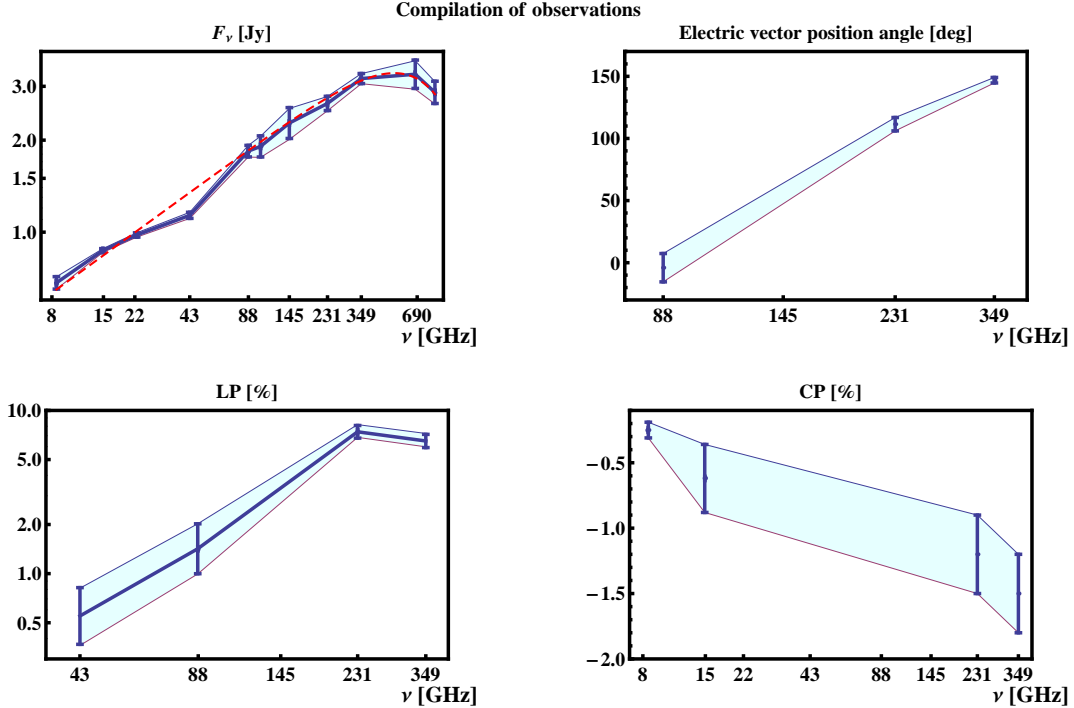}
\caption{Mean observed SEDs of specific flux $F_\nu$, linear
  polarization (LP) fraction, electric vector position angle (EVPA),
  and circular polarization (CP) fraction. The error bars show
  $1\sigma$ standard error of the mean. The dashed line on the $F_\nu$
  plot represents the analytic approximation $F_\nu({\rm
    Jy})=0.248\nu^{0.45}\exp(-(\nu/1100)^2)$ for frequency $\nu$ in
  GHz (not the simulated SED). As noted in Table~\ref{tab_obs}, the
  error is instrumental for CP at high frequencies and LP at $88$~GHz,
  whereas it is computed from a sample of observed quantities for
  flux, EVPA at all frequencies, and LP at high
  frequencies.}\label{fig_SED}
\end{figure*}

After the sample of fluxes, polarization fractions, and EVPAs
are found for each frequency group, we compute the mean and the standard
error. The summary of results is presented in Table~\ref{tab_obs}. CP fractions of $-1.2\%$
at $230$~GHz and $-1.5\%$ at $349$~GHz are based on SMA observations
by \citet{munoz} with the reported $\pm0.3\%$ instrumental
error. Note that standard errors in our total flux samples are
smaller than the error bars of prior observations
\citep{falcke98,yuan_data,broderick}, but still larger
compared to contemporary single-observation instrumental errors
\citep{marrone}. That is, we do not incorporate instrumental error in
our estimates of standard error of the mean fluxes or
$\LP$ and $\EVPA$ at $230$~GHz and $349$~GHz (even though
the instrumental error of $\LP$ at $88$~GHz is large). We do not
incorporate the source size measurements \citep{doeleman} in
calculating $\chi^2/\dof$, but we check that the best bet model is not
inconsistent with those observations.
Figure~\ref{fig_SED} shows a compilation of the mean quantities and their Gaussian standard
errors. The data are represented by both error bars and the
interpolated shaded area. A red dashed curve on the $F_\nu$
plot represents the analytic approximation
$F_\nu=0.248\nu^{0.45}\exp(-(\nu/1100)^2)$, where flux is in Jy and
frequency is in GHz.

\section{THREE-DIMENSIONAL GRMHD SIMULATIONS}
\label{s_simul}

Our radiative transfer calculations take the results of simulations of
accretion flows onto BHs as input.  These simulations are similar to
those in \citet{penna}.  Below, we review the methodology.

\subsection{Governing Equations}

We simulate radiatively inefficient accretion flows (RIAFs) onto
rotating BHs using a three-dimensional fully general relativistic code
(see \S~\ref{sec:nummethods}).  The BH is described by the Kerr
metric. We work with Heaviside-Lorentz units. Our five simulations
correspond to different choices of the dimensionless BH spin
parameter: $a_*=0, 0.5, 0.7, 0.9$, and $0.98$.  The self-gravity of
the RIAF is ignored.

The RIAF is a magnetized fluid, so we solve the GRMHD equations of
motion \citep{gammie}. Mass conservation gives:
\begin{equation}
\nabla_\mu (\rho u^\mu) = 0 ,
\end{equation}
where $\rho$ is the fluid frame rest-mass density, $u^\mu$ is the
contravariant 4-velocity, and $\nabla_\mu$ is the covariant
derivative. Energy-momentum conservation gives
\begin{equation}\label{emomeq}
\nabla_\mu  T^\mu_\nu = 0,
\end{equation}
where the stress energy tensor $T^\mu_\nu$ includes both matter and
electromagnetic terms,
\begin{equation}
T^\mu_\nu = (\rho + u_{\rm gas} + p_{\rm gas} + b^2) u^\mu u_\nu + (p_{\rm gas} + b^2/2)\delta^\mu_\nu - b^\mu b_\nu,
\end{equation}
where $u_{\rm gas}$ is the internal energy density and $p_{\rm gas}=(\Gamma-1)u_{\rm gas}$ is
the ideal gas pressure with $\Gamma=4/3$. Models with $\Gamma=5/3$
show minor differences compared to models with $\Gamma=4/3$
\citep{mg04,mignone}.  The contravariant fluid-frame magnetic 4-field
is given by $b^\mu$ and is related to the lab-frame 3-field $B^mu$ via $b^\mu
= B^\nu h^\mu_\nu/u^t$, where $h^\mu_\nu = u^\mu u_\nu +
\delta^\mu_\nu$ is a projection tensor and $\delta^\mu_\nu$ is the
Kronecker delta function \citep{gammie}.  We often employ $\bf b$
below, which is the orthonormal magnetic field vector in a comoving
locally flat reference frame \citep{penna}.  The magnetic energy
density ($u_b$) and magnetic pressure ($p_{\rm mag}$) are then given by
$u_{\rm mag}=p_{\rm mag}=b^\mu b_\mu/2 = b^2/2={\bf b}^2/2$.  Note that the angular
velocity of the gas is $\Omega=u^\phi/u^t$.

Magnetic flux conservation is given by the induction equation
\begin{equation}
\partial_t(\detg B^i) = -\partial_j[\detg(B^i v^j - B^j v^i)] ,
\end{equation}
where $v^i=u^i/u^t$, and $g={\rm Det}(g_{\mu\nu})$ is the determinant
of the metric.  No explicit resistivity or viscosity is included,
but we use a shock-capturing Godunov method that fully conserves energy.
So, all dissipation from shocks and numerical diffusivity
(e.g. in shear flows or current sheets) is fully captured,
as required to study RIAFs.

In \citet{penna}, we studied both RIAFs and geometrically thin
radiatively efficient disks. For the later case, a cooling term was
added to the energy-momentum equation (\ref{emomeq}) to describe
radiative losses and keep the disk thin.  The current set of models
are all RIAFs, so no cooling term is used.  Entropy generated by
viscous or resistive dissipation is advected along with the inflow
or transported out via convection or in a wind.

\subsection{Physical Models}

The initial mass distribution is an isentropic equilibrium torus
\citep{chak85a,chak85b,villiers} with pressure $p=K_0\rho^{4/3}$ for
$K_0=0.009$.  The torus inner edge is at $r_{\rm in}=20M$ and the maximum
density and pressure are at $R_{\rm max}=65M$.  We initialize the
solution so that $\rho=1$ at the pressure maximum.  As in
\citet{chak85a}, the angular velocity distribution of the initial
torus is a power law, where for the \citet{chak85a} $q$-parameter we
choose $q=1.65$ (At large radii $\Omega \sim (r/M)^{-q}$.).  The
thickness of the torus at the pressure maximum is then $|h/r|\sim 0.3$,
where
\begin{equation}\label{meantheta}
|h/r| \equiv
\frac{\int\int\int |\theta - \pi/2| \,\rho(r,\theta,\phi) dA_{\theta\phi}dt}
{\int\int\int \rho(r,\theta,\phi) dA_{\theta\phi}dt},
\end{equation}
where $dA_{\theta\phi}\equiv \sqrt{-g} d\theta d\phi$ is an area
element in the $\theta-\phi$ plane, and the integral over $dt$ is a
time average over the period when the disk is in a steady state (see
\S\ref{sec:convergence}).  A tenuous atmosphere fills the space
outside the torus.  It has the same polytropic equation of state as
the torus, $p=K_0 \rho^\Gamma$, with $\Gamma=4/3$, and an initial
rest-mass density of $\rho=10^{-6}(r/M)^{-3/2}$, corresponding to a
Bondi-like atmosphere. The torus is threaded with three loops of weak,
poloidal magnetic field: the initial gas-to-magnetic pressure ratio is
$\beta=p_{\rm gas,max}/p_{\rm mag,max}=100$, where $p_{\rm gas,max}$ and
$p_{mag,max}$ are the maximum values of the gas and magnetic pressure in
the torus.  This approach to normalizing the initial field is used in
many other studies \citep{gammie,mg04,m06a,mck07b,km07,penna}.

Recent GRMHD simulations of thick disks indicate that the results for
the disk (but not the wind-jet, which for us is less important) are
roughly independent of the initial field geometry
(\citealt{mck07a,mck07b,bhk08}, but see also \citealt{mtb12}).
The magnetic vector potential we use is given by
\begin{equation}\label{vectorpot}
A_{\phi,\rm N} \propto Q^2\sin\left(\frac{\log(r/S)}{\lambda_{\rm
field}/(2\pi r)}\right) \left[1 + 0.02({\rm ranc}-0.5)\right],
\end{equation}
with all other $A_\mu$ initially zero. This is the same $A_\mu$ as used in \citet{penna}.
We use $Q = (u_{\rm gas}/u_{\rm gas,max}
- 0.2) (r/M)^{3/4}$, and set $Q=0$ if either $r<S$ or $Q<0$. Here
$u_{g,\rm max}$ is the maximum value of the internal energy density in
the torus.  We choose $S=22M$ and $\lambda_{\rm field}/(2\pi r)=0.28$,
which gives initial poloidal loops that are roughly isotropic such
that they have roughly 1:1 aspect ratio in the poloidal plane.  The
form of the potential in equation (\ref{vectorpot}) ensures that each
additional field loop bundle has opposite polarity.  Perturbations are
introduced to excite the magneto-rotational instability (MRI).  The
second term on the right-hand-side (RHS) of equation \ref{vectorpot}
is a random perturbation: ${\rm ranc}$ is a random real number
generator for the domain $0$ to $1$.  Random perturbations are
introduced in the initial internal energy density in the same way,
with an amplitude of $10\%$. In \citet{penna}, it was found that
similar simulations with perturbations of $2\%$ and $10\%$ became
turbulent at about the same time, the magnetic field energy at that
time was negligibly different, and there was no evidence for
significant differences in any quantities during inflow equilibrium.

\subsection{Numerical Methods}
\label{sec:nummethods}

We perform simulations using a fully 3D version of HARM that uses a
conservative shock-capturing Godunov scheme
\citep{gammie,shafee,m06,noble06,mignone,tchekh07,mb09}.  We use
horizon-penetrating Kerr-Schild coordinates for the Kerr metric
\citep{gammie,mg04}, which avoids any issues with the coordinate
singularity in Boyer-Lindquist coordinates. The code uses uniform
internal coordinates $(t,x^{(1)},x^{(2)},x^{(3)})$ mapped to the
physical coordinates $(t,r,\theta,\phi)$.  The radial grid mapping is
\begin{equation}
r(x^{(1)}) = R_0 + \exp{(x^{(1)})} ,
\end{equation}
which spans from $R_{\rm in}=0.9 r_H$ to $R_{\rm out}=200M$, where
$r_H$ is the radius of the outer event horizon.  This just ensures the
grid never extends inside the inner horizon, in which case the
equations of motion would no longer be hyperbolic. The parameter
$R_0=0.3M$ controls the resolution near the horizon.  For the outer
radial boundary of the box, absorbing (outflow, no inflow allowed)
boundary conditions are used.

The $\theta$-grid mapping is
\begin{equation}
\theta(x^{(2)}) = \left[Y(2\xc-1) + (1-Y)(2\xc-1)^7 +1\right](\pi/2) ,
\end{equation}
where $x^{(2)}$ ranges from $0$ to $1$ (i.e. no cut-out at the poles)
and $Y=0.65$ is chosen to concentrate grid zones toward the equator.
Reflecting boundary conditions are used at the polar axes. The
$\phi$-grid mapping is given by $\phi(x^{(3)}) = 2\pi x^{(3)}$, such
that $x^{(3)}$ varies from $0$ to $1/2$ for a box with $\Delta\phi =
\pi$.  Periodic boundary conditions are used in the
$\phi$-direction. \citet{penna} considered various $\Delta\phi$ for
thin disks and found little difference in the results.  In all of
their tests, $\Delta\phi > 7 |h/r|$ and we remain above this limit as
well.  In what follows, spatial integrals are renormalized to refer to
the full $2\pi$ range in $\phi$, even if our computational box size is
limited in the $\phi$-direction.  For the purpose of radiative
transfer, we combine two identical regions of size $\Delta\phi = \pi$
preserving the orientation to obtain the span of full $2\pi$.

\subsection{Resolution and Spatial Convergence}

The resolution of the simulations is $N_r\times N_\theta\times
N_\phi=256 \times 64\times 32$.  This is the fiducial resolution of
\citet{penna}.  \citet{shafee} found this resolution to be sufficient
to obtain convergence compared to a similar $512\times128\times 32$
model.  In the vertical direction, we have about 7 grid cells per
density scale height.  Turbulence is powered by the MRI, which is
seeded by the vertical component of the magnetic field \citep{bh98}.
The characteristic length scale of the MRI is the wavelength of the
fastest growing mode:
\begin{equation}
\lambda_{\rm MRI}=2\pi\frac{v^z_A}{\Omega_0},
\end{equation}
where $v^z_A$ is the vertical component of the Alfv\'en speed.  We
find that the MRI is well-resolved in the midplane of disk both
initially and in the saturated state.

\citet{penna} studied convergence in $N_r$, $N_\theta$, and $N_\phi$
and found that models with $N_r=256$ or $N_r=512$, $N_\theta=64$ or
$N_\theta=128$, and $N_\phi=64$ or $N_\phi=32$ behaved similar for
disks with similar resolution across the disk.  Our resolution of the
MRI and prior convergence testing by \citet{penna} for
similarly-resolved disks justify our choice of grid resolution.  It is
currently not computationally feasible to perform a similar spin
parameter study at much higher resolutions, and future studies will
continue to explore whether such simulations are fully converged
\citep{hawley11,mtb12}.

\subsection{Ceiling Constraints}

During the simulation, the rest-mass density and internal energy
densities can become low beyond the corona, but the code remains
accurate and stable for a finite value of $b^2/\rho$, $b^2/u_{\rm gas}$, and
$u_{\rm gas}/\rho$ for any given resolution.  We enforce $b^2/\rho\lesssim
10$, $b^2/u_{\rm gas}\lesssim 100$, and $u_{\rm gas}/\rho\lesssim 10$ by injecting a
sufficient amount of mass or internal energy into a fixed zero angular
momentum observer (ZAMO) frame with 4-velocity
$u_\mu=\{-\alpha,0,0,0\}$, where $\alpha=1/\sqrt{-g^{tt}}$ is the
lapse.

\begin{figure}[!ht]\epsscale{1}
\plotone{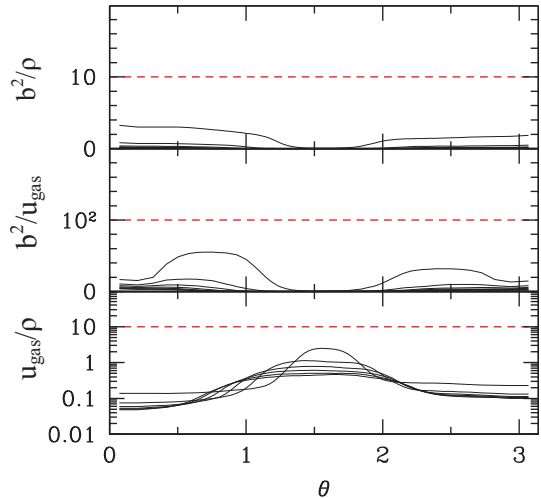}
\caption{Ratios of $b^2/\rho$, $b^2/u_{\rm gas}$, and $u_{\rm gas}/\rho$ versus
  $\theta$.  Black curves correspond to different radii in the flow;
  from top to bottom, $r=4, 6, 8, 10, 12$, and $14M$.  The data is
  time-averaged over the steady state period of the flow, from
  $t=14,000M$ to $20,000M$.  Numerical ceilings constrain the solution
  to lie below the dashed red lines, but we see that the solution does
  not approach these limits. }
\label{fig_floors}
\end{figure}

We have checked that the ceilings are rarely activated in the regions
of interest of the flow.  Figure \ref{fig_floors} shows the
constrained ratios, $b^2/\rho$, $b^2/u_{\rm gas}$, and $u_{\rm gas}/\rho$, as a
function of $\theta$ at six radii ($r=4, 6, 8, 10, 12$, and $14M$) for
the $a_*=0$ model.  The data has been time-averaged over the steady
state period from $t=14,000M$ to $20,000M$.  The ceiling constraints
are shown as dashed red lines.  The solution stays well away from the
ceilings.  Thus, the ceilings are sufficiently high.

\subsection{Approach to Steady State}
\label{sec:convergence}

\begin{figure}[!ht]\epsscale{1}
\plotone{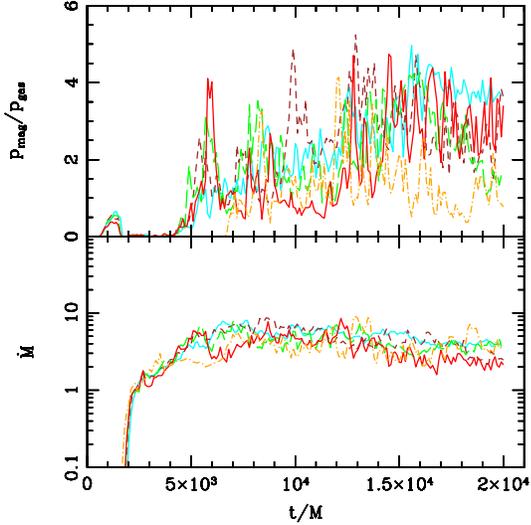}
\caption{Accretion rate and height- and $\phi-$averaged $\sigma=p_{\rm mag}/p_{\rm gas}=1/\beta$ versus
  time at the event horizon for all five models: $a_*=0$ (solid light
  cyan), $a_*=0.5$ (solid dark red), $a_*=0.7$ (long-dashed green),
  $a_*=0.9$ (short-dashed brown), and $a_*=0.98$ (dot-dashed
  orange).}\label{fig_convergence}
\end{figure}

We run the simulations from $t=0M$ to $t=20,000M$.  The accretion rate,
the height- and $\phi-$averaged plasma $\beta$, and other disk
parameters, fluctuate turbulently about their mean values.  The
simulation reaches a quasi-steady state, when the mean parameter value
are time-independent.  Figure \ref{fig_convergence} shows the
accretion rate and height- and $\phi-$averaged $1/\beta$ at the event
horizon as a function of time for all five models.  We take the period
from $t=14,000M$ to $t=20,000M$ to define steady state.

As shown in \citet{penna}, for disk models like the one considered,
the disk outside the innermost stable circular orbit (ISCO) behaves
like the $\alpha$-disk model with $\alpha\sim 0.1$ across disk
thicknesses of $h/r\sim 0.05-0.4$.  This allows one to accurately
infer the timescale for reaching ``inflow equilibrium,'' corresponding
to a quasi-steady flow across all quantities, at a given radius.  For
$h/r\sim 0.3$ by $t\sim 15,000M$-$20,000M$ (the simulation runs till
$20,000M$, but the initial $5,000M$ are transients not necessarily
associated with achieving inflow equilibrium for a simple viscous
disk), we use the results in Appendix B of \citet{penna} and find that
inflow equilibrium is achieved within a radius of $r\sim 25M$-$30M$
for models with $a_*\sim 1$ and $r\sim 35M$ for models with $a\sim 0$.
Even for a doubling of the viscous timescale, inflow equilibrium is
achieved by $r\sim 20M$-$25M$ depending upon the BH spin.  This
motivates using an analytical extension of the simulation solution for
radii beyond $r\sim 25M$ as described later in \S~\ref{subs_extent}.

\subsection{Evolved Disk Structure}

\begin{figure}[!ht]\epsscale{1}
\plotone{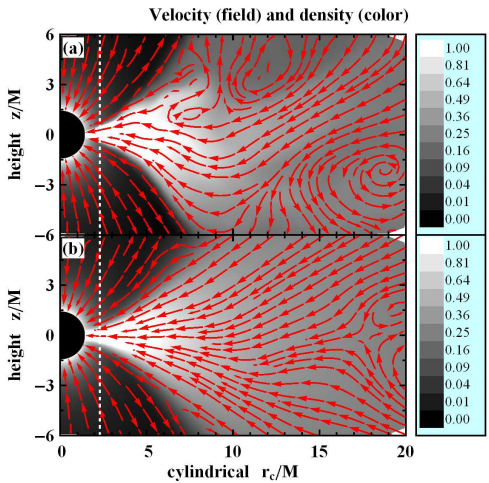}
\caption{Stream lines of velocity (red vectors) and number density
  $n_e$ (greyscale map) for spin $a_*=0.9$ at $\phi=0$ in the
  meridional plane($r_c$ as cylindrical radius): single time snapshot at $t=14,000M$ on the upper
  (a) panel and time average between $t=14,000M$ and $t=20,000M$ on the
  lower (b) panel. The corresponding calibration bars of $n_e$ are
  shown on the right. Number density is normalized by its maximum
  value, and the vectors show the poloidal velocity direction.}\label{fig_vfield}
\end{figure}

Figure~\ref{fig_vfield} shows matter stream lines as vectors and
number density $n_e$ as greyscale map. The large scale vortices
existing on a single time shot (panel a) almost disappear when
averaged over the duration $6,000M$ (panel b) from times $14,000M$ to $20,000M$. The
density is highest in the equatorial plane on average, but
deviations are present on the instantaneous map.  The ISCO does not
have any special significance: density and internal energy density
increase through ISCO towards the BH horizon.

\begin{figure}[!ht]\epsscale{1}
\plotone{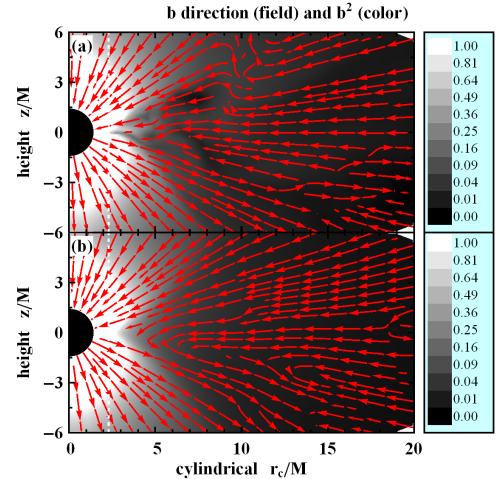}
\caption{Magnetic field lines (red vectors) and comoving
  electromagnetic energy density $\propto b^2$ (greyscale map) for
  spin $a_*=0.9$ at $\phi=0$ in the meridional plane($r_c$ as
  cylindrical radius): single time snapshot at $t=14,000M$ on the upper
  (a) panel and time average between $t=14,000M$ and $t=20,000M$ on the
  lower (b) panel. The corresponding calibration bars of comoving
  $b^2$ are shown on the right. Magnetic field energy density is
  normalized by its maximum value. The magnetic field lines illustrate
  only the direction of the field's poloidal component.}\label{fig_bfield}
\end{figure}

Figure~\ref{fig_bfield} shows magnetic field lines as vectors and
comoving electromagnetic energy density $\propto b^2$ as a greyscale
map.  The structure of magnetic field at early times remembers the
initial multi-loop field geometry \citep{penna}, but switches at late
times to a helical magnetic field structure resembling a
split-monopole in meridional projection. Such switching of magnetic
field structure suggests that the final helix with projected
split-monopole is a natural outcome of any vertical flux being dragged
into the BH (although the amount of magnetic flux threading the hole and disk
may be chosen by initial conditions as described in \citealt{mtb12}).
The magnetic field structure of a single snapshot (panel
a) looks similar to the structure of the linear average between
$14,000M$ and $20,000M$ (panel b).  The polar region of the flow has the
strongest magnetic field. The magnetic field lines on Figure~\ref{fig_bfield}
illustrate only the direction of the field's poloidal component. The toroidal magnetic field is stronger
above and below the midplane of the disk outside of ISCO. The toroidal field strength
is comparable to the poloidal field strength inside the ISCO
and near the disk midplane.

\section{DYNAMICAL MODEL BASED ON SIMULATIONS}

We now discuss extensions of the numerical simulations, which we need to
perform radiative transfer computations. We extend the simulations to
large radii and define the electron temperature.

\subsection{Extension to Large Radius}\label{subs_extent}

The flow is evolved in a quasi-steady state for $6,000M$ from $14,000M$
until $20,000M$, which corresponds to $8$ orbits at $r=25M$. The flow
is not sufficiently settled at larger radii. However, outside $25M$,
some Faraday rotation and emission might occur. So, we
extend the dynamical model to larger radii (i.e. $r>25M$) in a continuous way
and check (see Appendix~\ref{s_tests}) how variations of our large
radius prescriptions change the results of radiative transfer. The
outer radial boundary of radiative transfer is situated at $r=20,000M$. The profiles
of number density $n_e$, internal energy density $u_{\rm gas}$, magnetic field
$\bf b$, and velocity $\bf v$ are extended as power-laws until radius
$r=20,000M$. The power-law index for number density $\beta$ is obtained
by matching the known value $n_e=130{\rm cm}^{-3}$ at about
$1.5''\approx3\cdot10^5M$ \citep{baganoff} and the average $n_{e,\rm
  cut}$ value at $r=25M$ in the equatorial plane for each model. The
value of $\beta$ may be different for different models. The radial
flow velocity $v_r$ is then obtained from the continuity relation in
the equatorial plane $n_e v_r r^2=\rm const$. The power-law of
internal energy density $u_{\rm gas}$ is obtained in a similar way by matching
the values $T_e=T_p=1.5\cdot10^7$~K and $n_e=130{\rm cm}^{-3}$ at
distance $3\cdot10^5M$ \citep{baganoff,shcher_cond}. The meridional
physical velocity is extended as
$v_{\hat{\theta}}\propto(r/M)^{-3/2}$ and toroidal velocity as
$v_{\hat{\phi}}\propto(r/M)^{-1/2}$ to approximately match the power
law between $15M$ and $25M$, where the relationship
$v_{\hat{i}}\approx u^i\sqrt{g_{ii}}$ is used to connect the
4-velocity components with physical velocity components. All
components of comoving magnetic field are extended as $b_r, b_\theta,
b_\phi \propto (r/M)^{-1}$, which appears valid across a diverse set of
GRMHD models \citep{mtb12}. This power-law slope corresponds roughly
to equipartition of magnetic field energy density, since constant
fraction magnetic field is $b\propto \sqrt{n T_p}\propto (r/M)^{-1}$
for $n\propto (r/M)^{-1}$. Exploration of various extensions of the
magnetic field will be the topic of future studies.

After defining the extension power-laws for quantities in the
equatorial plane, we extend the quantities radially at arbitrary
$\theta$ and $\phi$ in a continuous way.  For example, for density at
arbitrary $\theta$ and $\phi$ and at $r>25M$ we have
\begin{equation}
n_e(r,\theta,\phi)=n_e(25M,\theta,\phi)\left(\frac{r}{25M}\right)^{-\beta},
\end{equation}
where $n_e(25M,\theta,\phi)$ is taken from the simulations. We
similarly extend other quantities. As shown in
Appendix~\ref{s_tests}, small variations in power-law indices of
number density and temperature have little influence on radiation
intensities and linear/circular polarization fluxes, but
variations of magnetic field slope can make a substantial difference.

\subsection{Electron Temperature}\label{sub_electron_t}

Neither the proton $T_p$ nor the electron $T_e$ temperature is given
directly by the simulations. However, it is crucial to know the
electron temperature $T_e$ to determine the emission. Our solution is
to split the total internal energy density $u_{\rm gas}$, given by the
simulations and their power-law extension, between the proton energy
and the electron energy. The energy balance states
\begin{equation}\label{eq_en_cons}
\frac{u_{\rm gas}}{\rho}\equiv \frac{u_{p,g}+u_{e,g}}{\rho}=c_p k_B T_p+c_e k_B T_e,
\end{equation}
where $c_p=3/2$ and $c_e\ge 3/2$ are the respective heat capacities,
$\rho$ is the rest-mass density, and $k_B$ is Boltzmann's
constant. The difference of temperatures $T_p-T_e$ is influenced by
three effects: equilibration by Coulomb collisions at large radii, the
difference in heating rates $f_p$ and $f_e$ of protons and electrons
operating at intermediate radii, and the difference in heat capacities
operating close to the BH.  Radiative cooling is ignored since,
according to \citet{sharma_heating}, the radiative efficiency of the
flow is negligible for realistic $\dot{M}\lesssim 10^{-7}M_\odot{\rm
  year}^{-1}$. The relevant effects can be summarized by the equation:
\begin{eqnarray}\label{eq_TeTp}
v_r\frac{d(T_p-T_e)}{dr}&=&-\nu_c(T_p-T_e)+\\
&+&\left(\frac{1}{c_p}\frac{f_p}{f_p+f_e}-\frac{1}{c_e'}\frac{f_e}{f_p+f_e} \right)v_r\frac{d(u_{\rm gas}/\rho)}{k_B dr},\nonumber
\end{eqnarray}
where
\begin{equation}
\nu_c=8.9\cdot10^{-11}\left(\frac{T_e}{3\cdot10^{10}}\right)^{-3/2}\frac{n_e}{10^7}
\end{equation}
is the non-relativistic temperature equilibration rate by collisions
\citep{shkarofsky}, all quantities being measured in CGS units. We
consider protons to always have non-relativistic heat capacity and
collisions to always obey the non-relativistic formula.  The
magnitudes of errors introduced by these simplifications are
negligible. The exact expressions for total electron heat capacity and
differential heat capacity are approximated as
\begin{eqnarray}
c_e&=&\frac{u_{e,g}/\rho}{k_B T_e}\approx\frac32\frac{0.7+2\theta_e}{0.7+\theta_e},\\
c_e'&=&\frac{d(u_{e,g}/\rho)}{k_B dT_e}\approx3-\frac{0.735}{(0.7+\theta_e)^2}
\end{eqnarray}
correspondingly, with the error $<1.3\%$, where
\begin{equation}\label{Te}
\theta_e=\frac{k_B T_e}{m_e c^2}
\end{equation}
is the dimensionless electron temperature.  It was recently shown
\citep{sharma_heating} that the ratio of heating rates in the
non-relativistic regime in a disk can be approximated as
\begin{equation}\label{heating}
\frac{f_e}{f_p}=C\sqrt{\frac{T_e}{T_p}}
\end{equation}
with coefficient $C$. This formula is adopted in the relativistic
regime as well, since no better prescription is
available. \citet{sharma_heating} found the value $C=0.33$ in
simulations, whereas we find $C=0.36-0.42$ for the best fit models
(see Table~\ref{tab_fit} and \S~\ref{s_result}).

\begin{figure}[!ht]\epsscale{1}
\plotone{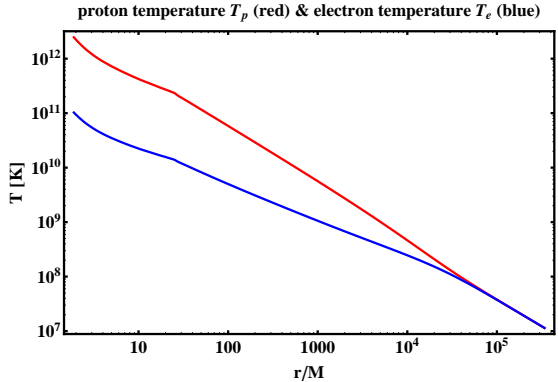}
\caption{Temperatures of protons $T_p$ (upper red line) and electrons
  $T_e$ (lower blue line) for the dynamical model with spin $a_*=0.5$
  giving the best fit to polarization observations (see
  Table~\ref{tab_fit} and \S~\ref{s_result}).}\label{fig_TpTe}
\end{figure}

The proton and electron temperatures are determined at each point in the following
way. We first take a single snapshot of a simulation with spin $a_*$
and extend the flow quantities to $r=20,000M$ (see \S~\ref{subs_extent}).
Then we compute azimuthal averages of radial velocity $v_r$, number
density $n_e$, and $u_{\rm gas}/\rho$ at the equatorial plane, extend them as
power laws to $r_{\rm out}=3\cdot10^5M$, and solve the equations
(\ref{eq_en_cons},\ref{eq_TeTp}) from $r_{\rm out}$ down to the inner
grid cell point.  Temperatures are set to $T_e=T_p=1.5\cdot10^7$~K at
$r_{\rm out}$ \citep{baganoff,shcher_cond}. On the next step we
compare the values of $u_{\rm gas}/\rho$ to the calculated $T_e$ and $T_p$ and
determine the functional dependence $T_e=T_e(u_{\rm gas}/\rho)$ and
$T_p=T_p(u_{\rm gas}/\rho)$. At each point of the simulation (including off
the equator), we draw temperatures from this correspondence. That is, GRMHD simulation directly provides
$u_{\rm gas}$ and $\rho$ at the equatorial plane, so the function $T_e=T_e(u_{\rm gas}/\rho)$ gives $T_e$ at each point in space.
Typical profiles of proton and electron temperatures are shown on
Figure~\ref{fig_TpTe}. Temperatures stay equal until $r\sim10^4M$ due
to collisions, despite different heating prescriptions. Within
$r=3\cdot10^3M$ the timescale of collisional equilibration becomes
relatively long and electrons become relativistic, thus $T_e$ deviates
down from $T_p$. The electron and proton temperature profiles
in the region $r<20,000M$ are used to conduct the radiative transfer.
For a given accretion rate, we find that there exists a unique dependence of the
ratio of temperatures $T_p/T_e$ (measured at $r=6M$ at the equator) upon the heating coefficient
$C$, so that we can use $T_p/T_e$ and $C$ interchangeably.

\section{GENERAL RELATIVISTIC POLARIZED RADIATIVE TRANSFER}\label{s_transf}
\subsection{Description of Radiative Transfer}
Now we convert the dynamical model of the accretion flow into a set
of observable quantities using polarized radiative transfer
\citep{broderick,shcher_huang}. We closely follow \citet{shcher_huang}
for the transfer technique. Similar to \citet{huangnew}, we define
the polarized basis in the picture plane, where one vector points
North, another vector points East, and the wavevector points towards
the observer. We parallel transport this basis in the direction of the
BH and do the radiative transfer along the ray in the opposite
direction towards the observer. At each point along the ray we go into the locally-flat
comoving frame, calculate the angles between the magnetic field and
basis vectors, and compute the Faraday conversion, Faraday rotation,
emissivities, and absorptivities.

Radiative transfer involves shooting a uniform grid of $P_N\times P_N$ geodesics
from the picture plane down to the black hole. The total polarized fluxes are computed
by integration of intensities along each ray backwards to the picture plane.
We found that $P_N=111$ is good enough to compute the spectrum (\citealt{dexter} used $P_N=150$).
For radiative transfer we employ all 3D data in each numerical simulation snapshot and, following \citet{moscibr_sim}, perform multilinear interpolation
in three dimensions for the quantities in between the grid points. We make no approximations in the use of spatial 3D data.
We self-consistently take into account the evolution of the numerical simulation as the light geodesics travel around the BH.
Since it is too time-consuming to look up simulation data over a long period of time, we only evolve the simulation between
$t-\Delta t$ and $t+\Delta t$ to get a spectrum at time $t+20,000M$.
The offset $20,000M$ appears, since the picture plane is located $20,000M$ away from the BH center.
The extension to the large radius outside $25M$, however, is not evolved with time. It is taken to be that of a single snapshot at time $t$.
The snapshot at times $t-\Delta t$ and $t+\Delta t$ are taken to represent the numerical simulations at earlier and later times, respectively.
We find that $\Delta t=60M$ is large enough to achieve accurate simulated spectra.
The total fluxes are found at regular time intervals within period of quasi-steady accretion from $14,000M$ till $20,000M$,
e.g. for $t=14,000M, 14,300M, ..., 19,700M, 20,000M$.
We compute $N_{\rm periods}=21$ spectra over the quasi-steady accretion phase and average them to find the mean simulated spectra.
To compute the polarized fluxes we take the integration domain in the picture plane to be a square with a side
\begin{equation}\label{region_size}
a[M]=16+2\left(\frac{600}{\nu[{\rm GHz}]}\right)^{1.5}
\end{equation} in the units of $r_{\rm g}\equiv M$, where frequency $\nu$ is in
GHz. This square is centered at the BH. The size based on Equation~(\ref{region_size}) is larger than the
photon orbit visible diameter $d_{\rm ph}\approx10.4M$ and
follows the intrinsic size dependence on frequency
\citep{shen,doeleman} at low frequencies. An important radiative
transfer parameter is the distance from the BH, where intensity
integration starts. The dependence of synchrotron emissivity on
temperature and magnetic field strength is so strong that it overwhelms
the sole effect of gravitational redshift close to the BH. We obtain
accurate results in the sub-mm for computation out from
$r_{\rm min}=1.01r_H$, where $r_H=M(1+\sqrt{1-a_*^2})$ is the horizon
radius. To quantify the needed accuracy of computations, we define a
quantity $\chi^2_H/\dof$ in Appendix~\ref{s_tests}.  We conduct
multiple tests of radiative transfer convergence for best fit
models at each spin.  In Appendix~\ref{s_tests}, we justify the chosen
values of radiative transfer parameters $P_N$, $\Delta t$, $N_{\rm
  periods}$, $r_{\rm min}$, etc.

Our calculation of plasma response is different from
\citet{shcher_huang}. They offered a way to find exact
emissivities, absorptivities, Faraday rotation, and conversion
coefficients for thermal and other isotropic particle
distributions. Here, for simplicity, we employ fitting formulas for
Faraday rotation and Faraday conversion and synchrotron approximation
for emissivities for a thermal plasma. We define
\begin{equation}\label{def_x}
X=\frac23\frac{\nu}{\nu_B\gamma^2\sin\theta_B},
\end{equation}
where $\theta_B$ is ${\bf k}$-${\bf b}$ angle, $\gamma$ is electron gamma factor, and $\nu_B=e b/(2\pi m_e c)$ is the cyclotron frequency. Then following \citet{legg,melrose_emis}, we write down emissivities in the $I$, $Q$, and $V$ modes as
\begin{eqnarray}\label{emissivities}
\varepsilon_I&=&\frac{\sqrt{3}}{2}\frac{e^2}{c}\nu_B\sin\theta_B\int^{+\infty}_1 d\gamma
N(\gamma)X\int^{+\infty}_X dz K_{5/3}(z),\nonumber\\
\varepsilon_Q&=&\frac{\sqrt{3}}{2}\frac{e^2}{c}\nu_B\sin\theta_B\int^{+\infty}_1 d\gamma
N(\gamma)X K_{2/3}(X),\\
\varepsilon_V&=&\frac{2}{\sqrt{3}}\frac{e^2}{c}\nu_B\cos\theta_B\int^{+\infty}_1 d\gamma\frac{N(\gamma)}{\gamma}\times \nonumber\\
&\times&\bigg[X K_{1/3}(X)+\int^{+\infty}_X dz K_{1/3}(z)\bigg].\nonumber
\end{eqnarray}
Here $K_z(x)$ is the Bessel function of the 2nd kind of order $z$. We
employed IEEE/IAU definitions of Stokes $Q$, $U$, and $V$
\citep{hamaker}, and we define counter-clockwise rotation of the electric field as seen by the observer
as corresponding to positive $V>0$ -- as also chosen in \citet{shcher_huang}.
So, the sign of the $V$ emissivity (Eq.~\ref{emissivities}) is opposite to the sign in
\citet{rybicki}. A variation of emissivity formulas
(\ref{def_x},\ref{emissivities}) exists: \citet{sazonov,pach} define
$X=2\nu/(3\nu_B(\gamma-1)^2\sin\theta_B)$, integrating over particle
energy instead of $\gamma$. This approximation appears to give
significantly larger errors at low particle energies.

Next, one needs to identify the accurate thermal particle distribution $N(\gamma)$.
Various $N(\gamma)$ correspond to various synchrotron approximations. The ultrarelativistic
thermal approximation \citep{pach,huangnew} has the simplest distribution
$N(\gamma)=\exp(-(\gamma-1)/\theta_e)(\gamma-1)^2/2/\theta_e^3$. However, the exact thermal distribution of
particles
\begin{equation}\label{distribution}
N(\gamma)=\gamma\sqrt{\gamma^2-1}\frac{\exp(-\gamma/\theta_e)}{\theta_e K_2(\theta_e^{-1})}
\end{equation}
allows for more precise computation of radiation.  Synchrotron
emissivities based on the equations (\ref{def_x},\ref{emissivities})
with the exact thermal distribution (\ref{distribution}) agree with
the exact cyclo-synchrotron emissivities $\varepsilon_I$,
$\varepsilon_Q$, and $\varepsilon_V$ \citep{leung,shcher_huang} to
within $2\%$ for typical dynamical models and frequencies
$>100$~GHz. Emissivities integrated over the ultrarelativistic thermal
distribution typically have $\sim10\%$ error.

Thermal absorptivities are found from emissivities
(Eq.~\ref{emissivities}) via Kirchhoff's law
\begin{equation}
\alpha_{I,Q,V}=\varepsilon_{I,Q,V}/B_\nu,
\end{equation}
where $B_\nu=2k_B T_e\nu^2/c^2$ is the source function for low photon
energies ($h\nu\ll k_B T_e$). Faraday rotation $\rho_V$ and Faraday
conversion $\rho_Q$ coefficients are taken from \citet{shcher_farad}:
\begin{eqnarray}
\rho_V=g(Z)\frac{2n_e e^2 \nu_B}{m_e c \nu^2}\frac{K_0(\theta_e^{-1})}{K_2(\theta_e^{-1})}\cos\theta,\\
\rho_Q=f(Z)\frac{n_e e^2 \nu^2_B}{m_e c \nu^3}\bigg[\frac{K_1(\theta_e^{-1})}{K_2(\theta_e^{-1})}+6\theta_e\bigg]\sin^2\theta.\nonumber
\end{eqnarray}
Here
\begin{equation}
Z=\theta_e\sqrt{\sqrt{2}\sin\theta\left(10^3\frac{\nu_B}{\nu}\right)}
\end{equation}
and
\begin{eqnarray}
g(Z)&=&1-0.11\ln(1+0.035Z),\nonumber\\
f(Z)&=&2.011\exp\left(-\frac{Z^{1.035}}{4.7}\right)-\\
&-&\cos\left(\frac{Z}{2}\right)\exp\left(-\frac{Z^{1.2}}{2.73}\right)-0.011
\exp\left(-\frac{Z}{47.2}\right)\nonumber
\end{eqnarray}
are the fitting formulas for deviations of $\rho_V$ and $\rho_Q$ from
analytic results for finite ratios of $\nu_B/\nu$. The deviation of
$f(Z)$ from $1$ is significant for the set of observed frequencies
$\nu$, temperatures $\theta_e$, and magnetic fields found in the
typical models of Sgr A*. These formulas constitute a good fit to the
exact result for the typical parameters of the dynamical model
\citep{shcher_farad}.

Polarized radiative transfer can take much longer to perform compared
to non-polarized radiative transfer when using an explicit integration
scheme to evolve the Stokes occupation numbers $N_Q$, $N_U$, and
$N_V$. Large Faraday rotation measure and Faraday conversion measure
lead to oscillations between occupation numbers. One of the solutions
is to use an implicit integration scheme, while another solution is to
perform a substitution of variables. In the simple case of Faraday
rotation leading to interchange of $N_Q$ and $N_U$, our choice of
variables is the amplitude of oscillations and the phase. Thus, the
cylindrical polarized coordinates arise as follows:
\begin{eqnarray}
N_Q&=&N_{QU}\cos\phi,\\
N_U&=&N_{QU}\sin\phi.\nonumber
\end{eqnarray}
Then, the amplitude $N_{QU}$ slowly changes along the ray and the angle
$\phi$ changes linearly, and this translates into a speed improvement. In
the presence of substantial Faraday conversion, the polarization
vector precesses along some axis on a Poincar\'e sphere, adding an
interchange of circularly and linearly polarized light. So, polar
polarized coordinates are more suitable in this case:
\begin{eqnarray}
N_Q&=&N_{\rm pol}\cos\phi\sin\psi,\nonumber\\
N_U&=&N_{\rm pol}\sin\phi\sin\psi,\\
N_V&=&N_{\rm pol}\cos\psi,\nonumber
\end{eqnarray}
where $N_{\rm pol}$ is the total polarized intensity,
$\phi$ angle changes are mainly due to Faraday rotation, and $\psi$ angle
changes are mainly due to Faraday conversion. The application of this technique
speeds up the code enormously at low frequencies of $\nu<100$~GHz.

\begin{table*}
\caption{Properties of the best fit models with different spins.}\label{tab_fit}
\begin{tabular}{| p{22mm} | p{19mm} | p{20mm} | p{19mm} | p{20mm} | p{22mm} | p{22mm}|}
\tableline\tableline
 Model & Inclination angle $\theta$, deg & Spin position angle PA, deg & Heating constant $C$ & Ratio $T_p/T_e$ at $6M$ & Electron $T_e$ at $6M$, K & Accretion rate $\dot{M}$, $M_\odot {\rm yr}^{-1}$ \\\tableline
spin $a_*=0$    & 42.0 & 171.0 & 0.42107 & 15.98 & $3.343\cdot10^{10}$ & $7.005\cdot10^{-8}$ \\\tableline
spin $a_*=0.5$  & 74.5 & 115.3 & 0.37012 & 20.14 & $3.087\cdot10^{10}$ & $4.594\cdot10^{-8}$ \\\tableline
spin $a_*=0.7$  & 64.5 & 84.7  & 0.37239 & 20.16 & $3.415\cdot10^{10}$ & $2.694\cdot10^{-8}$ \\\tableline
spin $a_*=0.9$  & 53.5 & 123.4 & 0.39849 & 18.16 & $4.055\cdot10^{10}$ & $1.402\cdot10^{-8}$ \\\tableline
spin $a_*=0.98$ & 57.2 & 120.3 & 0.41343 & 17.00 & $4.190\cdot10^{10}$ & $1.553\cdot10^{-8}$ \\\tableline
\hline
spin $a_*=0.5$ short period 1 & 70.0 & 79.3  & 0.38934 & 18.50 & $3.334\cdot10^{10}$ & $3.513\cdot10^{-8}$ \\\tableline
spin $a_*=0.5$ short period 2 & 72.8 & 113.1 & 0.40507 & 17.31 & $3.541\cdot10^{10}$ & $3.452\cdot10^{-8}$ \\\tableline
spin $a_*=0.5$ short period 3 & 73.4 & 57.4  & 0.37302 & 19.87 & $3.125\cdot10^{10}$ & $3.897\cdot10^{-8}$ \\\tableline
spin $a_*=0.5$ short period 4 & 74.4 & 115.4 & 0.36147 & 20.95 & $2.978\cdot10^{10}$ & $4.508\cdot10^{-8}$ \\\tableline
spin $a_*=0.5$ short period 5 & 71.9 & 95.7  & 0.37420 & 19.79 & $3.137\cdot10^{10}$ & $5.334\cdot10^{-8}$ \\\tableline
spin $a_*=0.5$ short period 6 & 76.4 & 116.7 & 0.38853 & 18.59 & $3.320\cdot10^{10}$ & $6.080\cdot10^{-8}$ \\\tableline
\hline
spin $a_*=0$ fast light    & 41.4 & 187.5 & 0.41929 & 16.09 & $3.322\cdot10^{10}$ & $7.044\cdot10^{-8}$ \\\tableline
spin $a_*=0.5$ fast light  & 72.7 & 105.9 & 0.39804 & 17.83 & $3.447\cdot10^{10}$ & $3.957\cdot10^{-8}$ \\\tableline
spin $a_*=0.7$ fast light  & 59.4 & 131.8 & 0.35708 & 21.62 & $3.204\cdot10^{10}$ & $2.966\cdot10^{-8}$ \\\tableline
spin $a_*=0.9$ fast light  & 53.3 & 123.3 & 0.40215 & 17.86 & $4.116\cdot10^{10}$ & $1.340\cdot10^{-8}$ \\\tableline
spin $a_*=0.98$ fast light & 57.7 & 119.6 & 0.41720 & 16.73 & $4.246\cdot10^{10}$ & $1.515\cdot10^{-8}$ \\\tableline
\end{tabular}
\tablecomments{Mean values are shown for ratio $T_p/T_e$, electron temperature $T_e$, and the accretion rate $\dot{M}$.
These are the simple means over all simulation snapshots, which were employed for radiative transfer in a particular model.
The values of $\chi^2/\dof$ ranges from $2$--$5$ across all models.}
\end{table*}

\begin{figure*}[!ht]\epsscale{0.9}
\plotone{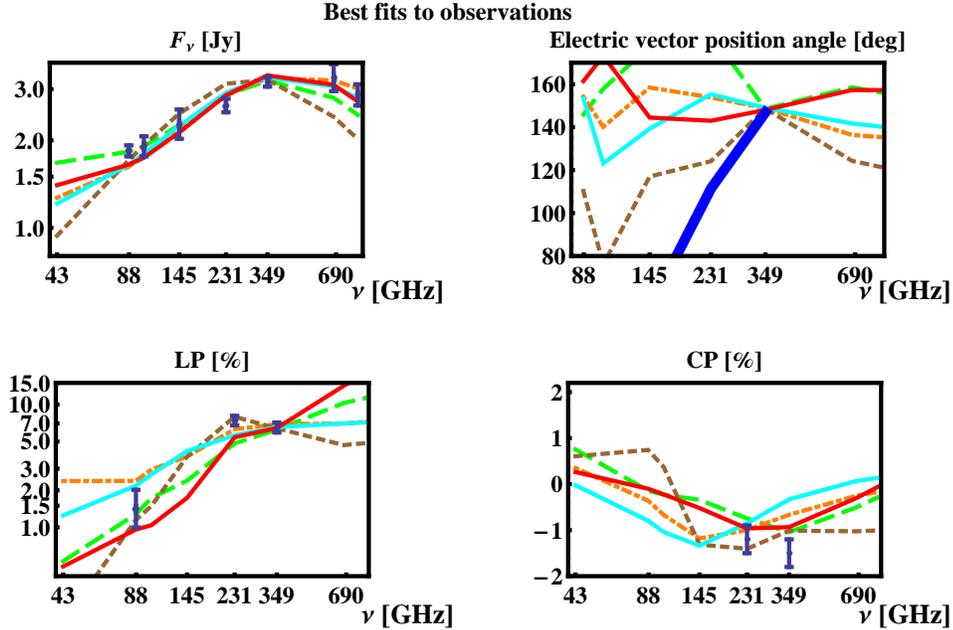}
\caption{Fits to the observed fluxes, LP and CP fractions by best
  models for each spin. The inclination angle $\theta$, accretion rate
  $\dot{M}$, ratio of temperatures $T_p/T_e$ were adjusted for each
  spin to minimize $\chi^2/\dof$. Fits to total flux $F_\nu$ are in the
  upper left panel, LP fraction in the lower left, and CP fraction in
  the lower right.  Shown are the best fit models with spin $a_*=0$
  (short-dashed brown), spin $a_*=0.5$ (solid dark red), spin
  $a_*=0.7$ (long-dashed green), spin $a_*=0.9$ (solid light cyan),
  and spin $a_*=0.98$ (dot-dashed orange).  The upper right panel shows the
  dependence of EVPA on frequency for the best models. Note,
  that EVPAs are not included into our fitting procedure. The
  thick blue curve represents observations. Simulated EVPA curves are
  arbitrarily shifted to approximate EVPA at $349$~GHz. The addition
  of an external (to the emitting region) Faraday rotation screen
  helps to fit ${\rm EVPA}(349~{\rm GHz})-{\rm EVPA}(230~{\rm
    GHz})$.}\label{fig_fitting}
\end{figure*}

\subsection{Search for the Best Fits}\label{s_analysis}

We define $\chi^2/\dof$ quantities to discriminate between models.
We define $\chi^2_F$ for fitting total fluxes as
\begin{equation}\label{chi_Fnu}
\chi^2_F=\sum_{i=1}^7 \frac{(F_{i,\rm sim}-F_{i, \rm obs})^2}{\sigma(F)^2},
\end{equation}
for the set of $7$ frequencies $\nu=88,102,145,230,349,680$, and
$857$~GHz, where $\sigma(F)$ are the errors of the means. We
incorporate LP fractions at $88, 230$, and $349$~GHz and CP fractions
at $230$ and $349$~GHz to obtain
\begin{eqnarray}\label{chi_FnuCPLP}
\chi^2=\chi^2_F&+&\sum_{i=1}^3 \frac{(\LP_{i,\rm sim}-\LP_{i, \rm obs})^2}{\sigma(\LP)^2}\nonumber\\
&+&\sum_{i=1}^2 \frac{(\CP_{i,\rm sim}-\CP_{i, \rm obs})^2}{\sigma(\CP)^2}.
\end{eqnarray}
Then we define $\dof$ (as degrees of freedom) to be $\dof_F=7-3=4$ for flux fitting
and $\dof=12-3=9$ for fitting all polarized data.
The quantity $\chi^2/\dof$ would be drawn from $\chi^2$ statistics if $\sigma$-s were the true observational errors
and if the observed fluxes were drawn from a Gaussian distribution.
However, for the purpose of the present work, we only employ $\chi^2/\dof$ as a measure of fitting the data.
That is, lower $\chi^2/\dof$ indicates better agreement with the data.
We do not attempt to ascribe any statistical meaning to the quantity $\chi^2/\dof$.

We explore models with $4$ parameters: spin $a_*$, inclination angle
$\theta$, accretion rate $\dot{M}$, and the ratio of proton to electron
temperature $T_p/T_e$ ($T_p/T_e$ is reported for radius $r=6M$). For the radiative transfer
calculations, the density from the simulations is scaled to give the desired
accretion rate.

\section{RESULTS}\label{s_result}

In previous sections, we described our compiled observations, GRMHD numerical simulations
of the flow structure, our method for obtaining the electron temperature,
and our method for polarized radiative transfer.
In this section, we discuss our results for accretion flow and BH parameters,
as guided by a minimization of $\chi^2/\dof$ for our model applied to the observations.

Figure~\ref{fig_fitting} shows best fits to observations by models
with five different spins. Inclination angle $\theta$, accretion rate
$\dot{M}$, and heating coefficient $C$ were adjusted to reach the lowest
$\chi^2/\dof$.  Fits to fluxes $F_\nu$ (upper left) are not
substantially different, although models with higher spins fit
better at high frequencies.  Larger deviations can be seen on $\LP$
(lower left) and $\CP$ (lower right) plots. Models with high spins
require lower accretion rate (i.e. density) to fit the flux spectrum. As a consequence,
they are not subject to Faraday depolarization, which leads to a
decrease of LP at low $\nu$, and the models end up having larger
linear polarization fractions at $88$~GHz. Not all models reproduce
the observed decrease of mean LP fraction between $230$~GHz and
$349$~GHz groups.  The discrepancies in fitting the CP fraction are also
large: all the lowest $\chi^2$ models give $|\CP|<1.5\%$ at $349$~GHz. The
best bet model with spin $a_*=0$ reproduces $\LP$ and $\CP$ fractions
well, but fails in fitting the total flux. Most solutions predict the
wrong sign of the $\EVPA(349~{\rm GHz})-\EVPA(230~{\rm GHz})$
difference, which could be fixed with stronger magnetic field
(e.g. as seen in models by \citealt{mtb12}) to yield stronger Faraday rotation.
In sum, crude agreement of simulated polarized spectra to the observed ones was achieved,
but the improved dynamical models may be needed for better fits.

We now isolate the physical effects responsible for the observed
polarized quantities for our best bet model with spin $a_*=0.5$
that has the lowest $\chi^2/\dof$ (see subsection~\ref{sec:model_param}).
\begin{figure}[!htbp]\epsscale{1}
\plotone{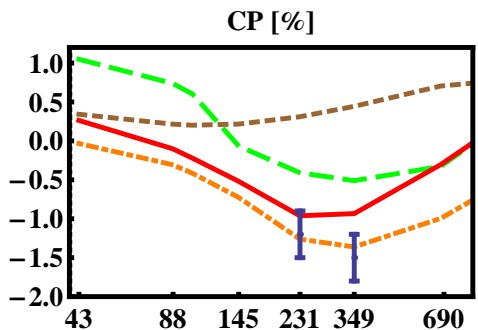}
\caption{Contributions of different effects to the CP fraction as a
  function of frequency for our best bet model with BH spin $a_*=0.5$. Shown are
  observations (blue error bars), the best bet model (solid red line),
  the same dynamical model computed with zero V emissivity
  ($\varepsilon_V=0$) in radiative transfer so that CP is produced by
  Faraday conversion (dot-dashed orange), the same model with zero
  Faraday conversion ($\rho_Q=0$) (short-dashed brown), and the same
  model with zero Faraday rotation ($\rho_V=0$) (long-dashed
  green). Emissivity in circular V mode contributes little to the
  observed CP, which is mainly due to Faraday conversion.}\label{fig_CP}
\end{figure}

\begin{figure*}[!ht]\epsscale{1}
\plotone{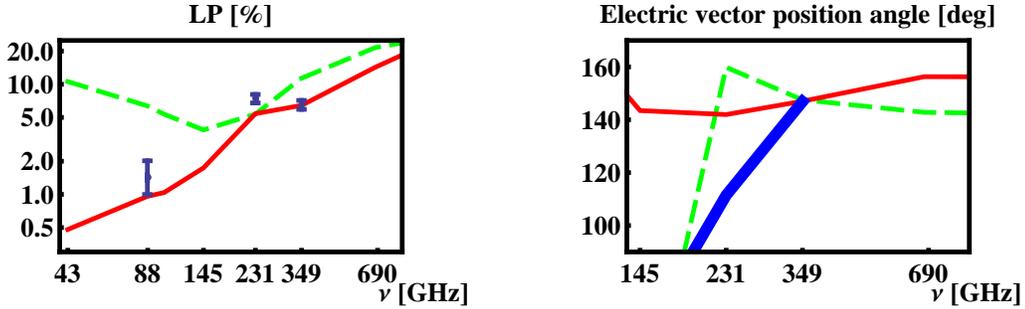}
\caption{Contributions of different effects to the LP fraction (on the
  left) and EVPA (on the right) as functions of frequency for
  the best bet model with spin $a_*=0.5$. Shown are observations (blue
  error bars and thick blue line), the best bet model (solid red
  line), and the same dynamical model computed with zero Faraday rotation
  ($\rho_V=0$) in radiative transfer (long-dashed green). Beam
  depolarization is weak: if Faraday rotation is absent, then LP
  stays high at low frequencies.  Even when the Faraday rotation is
  set to zero, the EVPA depends on frequency due to varying
  intrinsic emission EVPA. Faraday rotation in the best bet model
  is too weak to reproduce EVPA observations, so stronger magnetic
  fields or more magnetic flux near the black hole than in the simulations
  may be required.}\label{fig_LP}
\end{figure*}

There are several radiative transfer effects that contribute similarly
to the polarized fluxes. Let us consider the production of
circular polarization in the flow. Figure~\ref{fig_CP} shows the
consequence of switching off each physical effect for our
best bet model with spin $a_*=0.5$. The solid red curve is the result with
all physics on. The dot-dashed orange line below is for
zero circular emissivity having $\varepsilon_V=0$. The brown dashed line
corresponds to zero Faraday conversion ($\rho_Q=0$). Switching off
$\varepsilon_V$ emissivity leads to a minor correction, whereas
setting Faraday conversion to zero results in CP of the opposite sign with several
times smaller absolute value. Most of the CP in this model is produced by
Faraday conversion.  It would be incorrect, however, to think that the
simple linear to circular conversion explains the observed CP. The
dashed green line in Figure~\ref{fig_CP} shows the CP fraction, when
Faraday rotation is switched off ($\rho_V=0$). The effect of Faraday
rotation is insignificant at $\nu>350$~GHz, but the rotation of the
plane of linear polarization simultaneous with conversion between
linear and circular polarizations produces a unique effect at lower
$\nu$. This is the so-called ``rotation-induced conversion''
\citep{homan}. Sign oscillations of $V$ with frequency do not
happen when the Faraday rotation is on, but they do happen when
$\rho_V=0$. For the best fit model it is the rotation-induced Faraday rotation,
which is responsible for the most of circularly polarized light.

In Figure~\ref{fig_LP} we illustrate the influence of Faraday rotation
on LP fraction (left panel) and EVPA (right panel). The solid
curves are produced with all physics on for our best bet model with spin
$a_*=0.5$. The green dashed lines are computed when switching off
Faraday rotation ($\rho_V=0$).  The Faraday rotation is small at high
frequencies and $\LP$ curves look similar at $\nu>200$~GHz. As the
rotation of polarization plane is much stronger at low $\nu$, a
significant phase shift accumulates between different rays at the low
end of the spectrum and cancellations of LP become strong at
$\nu<150$~GHz. This illustrates the effect of Faraday depolarization
\citep{bower99a}. In the absence of Faraday rotation, the dependence of
EVPA on frequency is not constant: the variations of intrinsic emitted
EVPA are significant. Thus, the change of EVPA with $\nu$ should not
always be ascribed to the effect of Faraday rotation.  The positive
observed slope of EVPA with $\nu$ at high $\nu$, acquired due to
negative Faraday rotation measure ($RM<0$), is comparable to the slope
of intrinsic emitted EVPA.

\begin{figure}[!ht]\epsscale{1}
\plotone{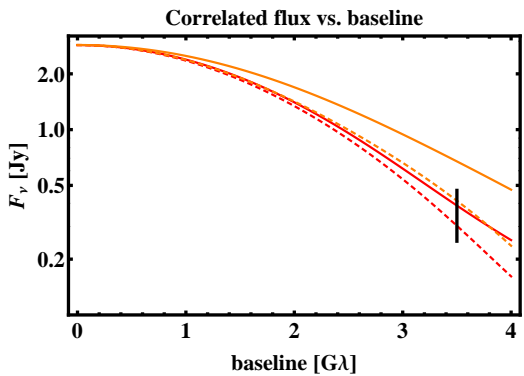}
\caption{Correlated flux as a function of baseline at $230$~GHz
  normalized to the averaged observed flux at $2.82$~Jy for the best fit models
  with spin $a_*=0.5$ (darker red lines) and $a_*=0.98$ (lighter orange lines).
   The upper solid lines show the smallest size (largest
  correlated flux) over all position angles of BH spin axis, and the lower dashed
  lines show the largest size (smallest correlated flux) over all
  position angles. An observational results presented in \citet{doeleman} with
  $3\sigma$ error bars at baseline $3.5$~G$\lambda$ is depicted as a vertical black bar for
  comparison. The size in our best bet model with spin $a_*=0.5$ is consistent with
  observations, whereas the best fit model with spin $a_*=0.98$ has larger correlated flux,
  so that the size of the shadow is slightly under-predicted.}\label{fig_size}
\end{figure}

There is an alternate way to test dynamical models against
observations. The intrinsic image size was recently measured
\citep{doeleman} with the VLBI technique. The measured correlated flux
at $230$~GHz was $F_{\rm corr}\approx0.35$~Jy at $3.5~{\rm G}\lambda$
SMT-JCMT baseline. Similar values of correlated flux were observed
later by the same group \citep{fish10}.  We plot this correlated flux
with $3\sigma$ error bar in Figure~\ref{fig_size} and compare it to
simulated correlated fluxes. To simulate the correlated flux we follow
\citet{fish} and employ a Gaussian interstellar scattering ellipse
with half-widths at half-maximum $7.0\times3.8{\rm G}\lambda$ with
position angle $170^\circ$ East of North.  The correlated fluxes for
the best fit models with spin $a_*=0.5$ (darker red lines) and
$a_*=0.98$ (lighter orange lines) are shown.  We vary the
position angle (PA) of the BH spin axis, and plot correlated flux curves with
the largest (upper solid lines) and the smallest (lower dashed lines)
correlated flux at $3.5{\rm G}\lambda$.  Since we do not fit EVPA
directly, models with different PA have the same $\chi^2/\dof$. The
size in our best bet model with spin $a_*=0.5$ is consistent with
observations, whereas the best fit model with spin $a_*=0.98$ has
larger correlated flux, so that the size of the shadow is slightly
under-predicted. The simulated source size is in crude agrement to the observed one.

Table~\ref{tab_fit} summarizes the properties of several best fit
models.  Rows $1-5$ show the model parameters for best fits with
spins from $a_*=0$ to $a_*=0.98$.  The simulated spectra are computed
every $300M$ from $t=14,000M$ till $t=20,000M$ for $\Delta t=60M$.
Then $N_{\rm period}=21$ spectra are averaged to compare to
observations.  The rows $6-11$ show the model parameters for models
with spin $a_*=0.5$ for spectra averaged over shorter periods.  That
is, $N_{\rm period}=21$ spectra are computed from $t=14,000M$ till
$t=15,000M$ for the $1$-st short period, while the second short period covers
the time interval from $t=15,000M$ till $t=16,000M$, etc.  When
comparing the best fit models with spin $a_*=0.5$ computed over
different simulation periods, we find variations in inclination angle
$\Delta \theta=3^\circ$ from the mean, the electron temperature
$\Delta T_e/T_e=10\%$, and the accretion rate $\Delta
\dot{M}/\dot{M}=30\%$. The spin position angle varies by as much as
$\Delta {\rm PA}=30^\circ$.

The last $5$ rows in Table~\ref{tab_fit} show the model parameters for best fits within the ``fast light''
approximation.  In this approximation, simulated spectra are computed
over single frozen snapshots, e.g. for $\Delta t=0$.  When the fast
light approximation is used instead of the correct simultaneous
evolution of photon field and MHD, the models with spins
$a_*=0;0.9;0.98$ produce almost identical best fits with variations
$\Delta \theta<0.6^\circ$, $\Delta T_e/T_e<1.5\%$, and $\Delta
\dot{M}/\dot{M}<5\%$. However, the models with $a_*=0.5; 0.7$ settle
to different $\chi^2/\dof$ minima with larger changes in quantities:
$\Delta \theta=5^\circ$, $\Delta T_e/T_e=10\%$, $\Delta
\dot{M}/\dot{M}=10\%$. These variations are still smaller than variations between models with different spins.
Switching to the fast light approximation
results in significant changes $\Delta\chi^2/\dof\sim1$ between the
best fit models for the same spins, which emphasizes the
importance of precise radiation transfer calculations.

\subsection{Model Parameters}\label{sec:model_param}

We now discuss the estimated parameters obtained for the best fit models.
The best bet model with spin $a_*=0.5$ has inclination angle $\theta=74.5^\circ$,
mean accretion rate $\dot{M}=4.6\times10^{-8}M_\odot{\rm year}^{-1}$,
ratio of temperatures $T_p/T_e=20.1$ at $r=6M$, which gives
$T_e=3.1\cdot10^{10}$~K at $r=6M$ in the equatorial
plane. The best fit models with other spins give the inclination angles:
$\theta=42^\circ 64.5^\circ, 53.5^\circ, 57.2^\circ$ at $a_* = 0; 0.7;
0.9; 0.98$, respectively. Thus, the inclination angle for the 5 models
lies within $\theta=42^\circ-75^\circ$. Our modeling favors neither
edge-on nor face-on orientations. The electron temperature $T_e$ at
$r=6M$ is surprisingly uniform over a set of
best fit models. All $5$ best fit models with spins from $a_*=0$
to $a_*=0.98$ presented in Table~\ref{tab_fit} have electron
temperature within the tight range
\begin{equation}
T_e = (3.0-4.2)\times 10^{10} {\rm K}.
\end{equation}
The accretion rate depends strongly on spin.  The model with
spin $a_* = 0$ has an accretion rate $\dot{M}=7.0\times10^8{M_\odot
  \rm year}^{-1}$, which is 5 times larger than the accretion
rate $\dot{M}=1.4\times10^8{M_\odot \rm year}^{-1}$ for the model with
spin $a_* =0.9$. Higher spin values give lower accretion rates. A
natural outcome of fitting polarized spectrum is the PA of the BH spin axis.
Similar to \citet{huangnew}, we rely on the observed intrinsic
EVPA$\approx111.5^\circ$ at $230$~GHz and EVPA$\approx146.9^\circ$ at
$349$~GHz (see \S~\ref{s_obs}). For the model to fit the difference in
EVPA, we add a Faraday rotation screen far from the BH with constant
rotation measure (RM). Then we compute the required RM and the
intrinsic PA to fit the simulated EVPAs at $230$ and $349$~GHz.  The
best bet model with $a_* = 0.5$ gives ${\rm PA} = 115.3^\circ$
East of North, whereas the next best fit model with spin $a_* = 0.98$
requires ${\rm PA}= 120.3^\circ$.  However, PA is different by
$90^\circ$ between the models with spin $a_* = 0$ and $a_* = 0.7$,
which indicates that PA can lie within a wide range.
In sum, some parameters, such as $T_e$, are estimated to be in narrow ranges,
while only order of magnitude estimates are available for other parameters, such as $\dot{M}$.

\begin{figure*}[!ht]\epsscale{1}
\plotone{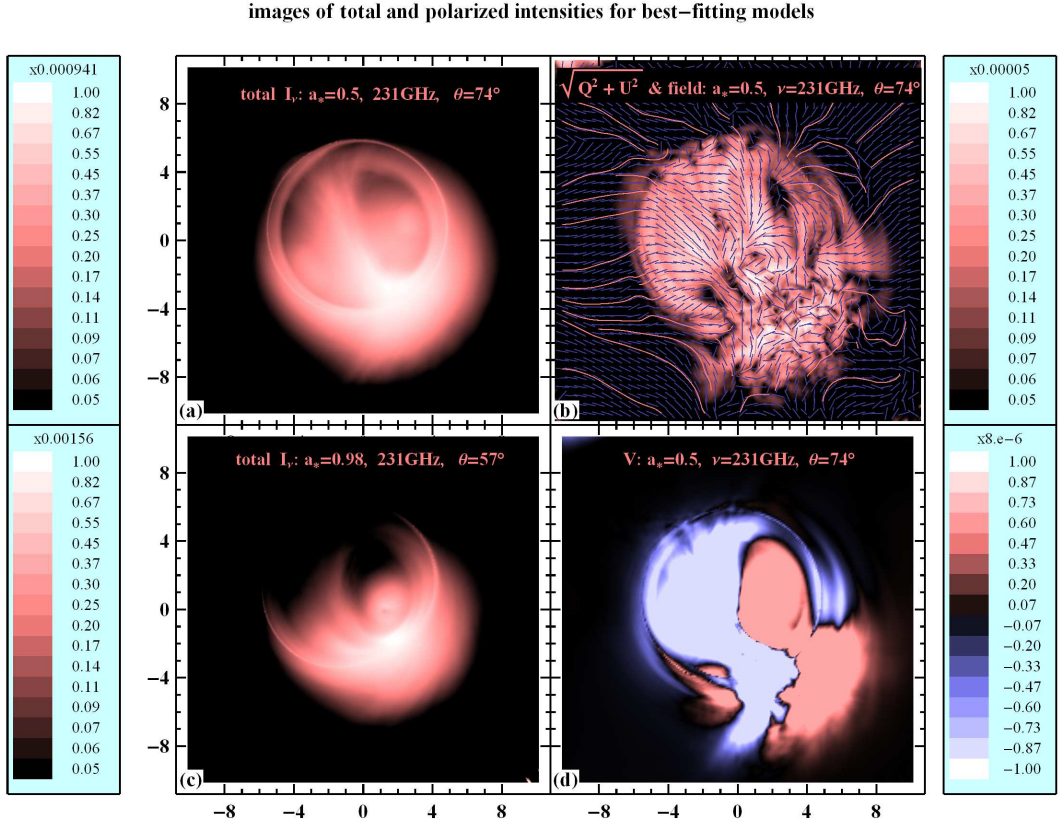}
\caption{Images of polarized intensities for the best fit models: total intensity for spin $a_*=0.98$ model (lower left);
intensities for $a_*=0.5$ model: total intensity (upper left), linear polarized intensity and streamlines along EVPA (upper right),
and circular polarized intensity (lower right). Distances are in the units of BH mass $M$.
Images are rotated in the picture plane to correspond to the best spin PA: ${\rm PA}=115.3^\circ$ for the $a_*=0.5$ model
and ${\rm PA}=120.3^\circ$ for the $a_*=0.98$ model. Individual calibration bars are on the sides of corresponding plots.
The ill-defined polar region does not contribute significantly to the emission.}\label{fig_images}
\end{figure*}

With the estimated orientation of the BH spin axis, we can plot an image of
average radiation intensity from near the event
horizon. Figure~\ref{fig_images} shows images of total intensity
$I_\nu$ for the best bet model with spin $a_*=0.5$ (upper left
panel), the best fit for spin $a_*=0.98$ (lower left panel), and LP
intensity and CP intensity plots for the best bet model with $a_*=0.5$ (upper
right and lower right panels, correspondingly). The LP average intensity
plot was made by averaging $U$ and $Q$ intensities separately and then
finding the total LP fraction and EVPA. Blue (predominant) color
on the CP plot depicts the regions with negative CP intensity and red
(scarce) color depicts the regions with positive CP intensity.  The
total $V$ flux from this solution is negative ($V<0$). The streamlines
on the LP plot are aligned with EVPA direction at each point. The spin
axis is rotated by ${\rm PA}=115.3^\circ$ East of North for the best bet model with spin
$a_*=0.5$ and by ${\rm PA}=120.3^\circ$ for the best fit model with spin
$a_*=0.98$. The spin axis is inclined at $\theta$
to the line of sight, so that the either right (West) or left (East)
portions of the flow are closer to the observer. The color schemes on
all plots are nonlinear with corresponding calibration bars plotted on
the sides. The numbers at the top of calibration bars denote
normalizations.

\section{DISCUSSION AND CONCLUSIONS}\label{s_discus}

Let us compare our results with estimates of Sgr A* accretion flow and
BH parameters made by other researchers.

Two separate searches for spin based on GRMHD numerical simulations have been reported so far:
\citet{moscibr_sim} and \citet{dexter2010}. \citet{moscibr_sim} considered
the set of spins from $a_*=0.5$ to $a_*=0.98$ for 2D GRMHD
simulations, then fitted the X-Ray flux, the $230$~GHz flux, and the flux slope at $230$~GHz.
They found at least one model for each spin is crudely consistent with the
observations (see their table~3), and their best bet model has
$a_*=0.9$. \citet{dexter2010} focused on a set of 3D GRMHD, then fitted the
$230$~GHz flux and size estimates, and they provided a table of spin
probabilities with $a_*=0.9$ having the highest $P(a)$. When we fitted the
spectrum and LP/CP fractions, the model with $a_*=0.5$ has the lowest $\chi^2/\dof$.
As for these two groups, we are also unable to provide a statistically significant constraint on $a_*$.
Other spin estimates have been based on analytic
models. \citet{broderick,broderick10} favor $a_*=0$ solutions, while
\citet{huangspin} favor $a_*<0.9$ (although they do not explore their
full model parameter space).

Another poorly constrained quantity is the
mass accretion rate. Our estimate $\dot{M}_{\rm
  est}=(1.4-7.0)\cdot10^{-8}M_\odot{\rm year}^{-1}$ is
broad. Acceptable models in \citet{moscibr_sim} give $\dot{M}$ from
$0.9\cdot10^{-8}M_\odot{\rm year}^{-1}$ to $12\cdot10^{-8}M_\odot{\rm
  year}^{-1}$, which agrees with our range. \citet{dexter2010}
reported $90\%$ confidence interval of $\dot{M}$ for spin $a_*=0.9$
solutions, while incorporating flow size in $\chi^2$ analysis. Our
estimates have somewhat higher accretion rates than the range
$\dot{M}=5^{+15}_{-2}\times10^{-9}M_\odot{\rm year}^{-1}$ ($90\%$) in
\citet{dexter2010}, because models with lower spin naturally need higher
$\dot{M}$ to fit the data. Note that \citet{dexter} found even lower
accretion rate $\dot{M}(a_*=0.9)=(1.0-2.3)\times10^{-9}M_\odot{\rm
  year}^{-1}$ when they assumed equality of proton and electron
temperatures ($T_p=T_e$).

In addition to spin and accretion rate, we can try to estimate the inclination
angle $\theta$ and electron temperature $T_e$ ($T_e$ is reported at $r=6M$ in the equatorial plane).
Our range is $\theta_{\rm est}=42^\circ-75^\circ$, which agrees with estimates by
other groups.  \citet{broderick,dexter2010} reported $\theta \sim 50^\circ$.
\citet{huangnew} and \citet{huangspin} favor
slightly lower $\theta\sim 40^\circ$ and $45^\circ$, respectively, but they have large error
bars. To estimate $T_e$, \citet{moscibr_sim} and \citet{dexter2010} use a
constant $T_p/T_e$, whereas \citet{huangnew} and the present work
calculated the profile of $T_e$. In all models, $T_e$ is a shallow
function of radius, which made \citet{dexter2010} estimate a
``common'' $T_e=(5.4\pm3.0)\times10^{10}~{\rm K}$
(calculated at some distance from the BH center).
We measure $T_e$ at $r=6M$, and we obtain a narrower range
(likely owing to fitting of polarized observations)
of $T_e=(3.0-4.2)\times10^{10}$~K.

One can use two types of observations to estimate the BH spin axis position angle:
the $230$~GHz correlated flux and the EVPA. Using the correlated flux gave
\citet{broderick} and \citet{dexter2010} a result of ${\rm
  PA}=(-70^\circ)$--$(-20^\circ)=(110^\circ)$--$(160^\circ)$.
Using the EVPA data has given slightly different results: \citet{meyer}
predicts the range ${\rm PA}=60^\circ-108^\circ$, whereas Huang gets
either ${\rm PA}\approx115^\circ$ \citep{huangspin} or ${\rm
  PA}\approx140^\circ$ \citep{huangnew}. Our values of PA are within
the range $85^\circ-171^\circ$, which is consistent with predictions in \citet{meyer} and with
estimates based upon the observed correlated flux.
The size of the flow may depend
substantially on the luminosity state \citep{broderick} or the presence of
non-thermal structures, spiral waves, and other features. In some
astrophysical sources, PA is directly known from spatially resolved
jets, and Sgr A* may be one of such sources. A tentative jet feature
was revealed in X-rays by \citet{muno_diff} in their fig.~8 showing ${\rm
  PA}_{\rm jet}=120^\circ$. This value is close to ${\rm
  PA}=115.3^\circ$ or ${\rm PA}=120.3^\circ$ for the best fit
models with spins $a_*=0.5$ and $a_*=0.98$, respectively.

Besides the estimates of accretion rate and flow properties based on
the inner flow, there exist estimates based on the outer
flow. \citet{shcher_cond} constructed an inflow-outflow model with
conduction and stellar winds with radiation matching the X-ray surface brightness
profile observed by \textit{Chandra}. Their model\footnote{Note that
  gravitational radius is defined as $r_{\rm g}=2M$ in
  \citet{shcher_cond}.} had an accretion
rate $\dot{M}=6\cdot10^{-8}M_\odot{\rm year}^{-1}$ and electron
temperature $T_e=3.6\times10^{10}~{\rm K}$ at $r=6M$, which is consistent
with present results. \citet{shcher_cond} constrained the density in the
outer-radial flow from X-ray observations, while the present work constrains the density
in the inner-radial flow from sub-mm observations. The density profile is
then found to be
\begin{equation}
\rho\propto r^{-\beta}, \quad \beta=0.80-0.90.
\end{equation}
The density power-law index $\beta$ lies between
$\beta=1.5$ for ADAF flow \citep{narayan} and $\beta=0.5$ for the
convection-dominated accretion flow
\citep{narayan00,quataert_cdaf}. However, the modification of the
power-law index from the steep ADAF profile is likely due to
conduction for Sgr A*, not convection \citep{shcher_cond}.
Newer GRMHD simulations of radially-extended disks show
a comparable power-law index for density \citep{mtb12}.

Our dynamical model has limitations and relies on several approximations.
More convergence testing, like done in \citet{mtb12}, is required to ensure the 3D GRMHD simulation results are reliable.
The amount of initial magnetic flux and the field geometry might have a pronounced effect on simulation results.
For example, magnetically choked accretion flows (MCAF) \citep{igumenshchev,mtb12} may
have more desirable properties (such as larger Faraday rotation as discussed related to Figure~\ref{fig_LP})
for SgrA* compared to MRI-dominated disks described in the present work.
The dependence of the estimated accretion flow and the BH parameters
on the simulation type and the initial setup should be carefully explored in future works.
The polarization is expected to be able to best highlight changes in the magnetic
field geometry and strength, and so our work is an important stepping stone to distinguish whether SgrA*
is a classical MRI-dominated disk or an MCAF.

The limited dynamical range of our simulations leads to another caveat.
We fix electron density $n_e$ and temperatures $T_p$ and $T_e$ in the outer flow and extend them down to the event horizon.
The slopes of these quantities break at $25M$ radius, where the power law radial extrapolation starts.
Thus, the density and temperature slopes in the inner flow may need to be determined more self-consistently.
Future simulations will need to cover a larger
range of radii and plasma physics effects, such as conduction \citep{johnson,sharma_cond,shcher_cond}.
Simulations with larger outer radial boundaries that are run for longer will also help to fit
the Faraday rotation, which happens for the present models partially
outside of the simulated domain. A proper simulation of the polar
region of the flow may be important as well. At present, we
artificially limit the lowest density and highest temperature there. If we do not,
then numerical artifacts associated with excessive numerical
dissipation and heating appear (similar to those in \citealt{moscibr_sim}).

We found the lowest $\chi^2/\dof$ for the model with spin $a_*=0.5$,
whereas other groups found $a_*=0$ and $a_*=0.9$ to provide the best fits in their modeling.
So, there still appears to be no reliable estimate of BH spin for SgrA*. One common shortcoming of recent papers
is the use of thermal electron distribution. If non-thermal electrons provide most of the energy
for the sub-mm peak, then this would invalidate all prior spin estimates
\citep{shcher_huang}.

The radiative transfer we performed has its shortcomings. The emissivities in our
special synchrotron approximation provide, e.g., $2\%$ agreement with
exact emissivities \citep{leung,shcher_huang} for $b=20$~G,
$\theta_B=1$~rad, $T_e=6.9\cdot10^9$~K, and observed frequency
$\nu=100$~GHz. Agreement is better for larger $T_e$. Non-polarized
radiative transfer methods \citep{moscibr_sim,dexter2010}
have an intrinsic error that is comparable with our polarized radiative transfer
method for the same total emissivity $\varepsilon_I$, but the error is still
$1-5\%$.

There are other unaccounted sources of error. The mass of the BH
in the Galactic Center is known to within $10\%$ \citep{ghez,gillessen} and the
distance is known to $5\%$. We do not expect these uncertainties to
lead to significant changes in our predictions. A shift to slightly
lower spin may be able to mimic the effect of smaller BH or a BH at
larger distance.

An improvement in observations can lead to further insights on the flow and black hole parameters. For example,
the detailed comparison of flux, LP,
and CP curves in Figure~\ref{fig_fitting} shows that the models with
different spins have discrepancies at frequencies not yet
probed by observations. In particular, the CP fractions at
$88$~GHz and $690$~GHz are different. The EVPA data need improvement as well.
EVPA observations are available at $230$~GHz and $349$~GHz, but these frequencies are affected by Faraday rotation.
The observations at higher frequency, where the Faraday rotation effect is weaker,
should provide a better estimate of BH spin axis PA.
Another important quantity, LP at $88$~GHz, has a largely unknown value.
Its observations are reported in $2$ papers. Variations
in simulated $\LP(88{\rm GHz})$ are large between the best fit models (see
Figure~\ref{fig_fitting}). Refinement of the observed mean $\LP(88{\rm
  GHz})$ could potentially help discriminate between different spins.
A measurement of the emitting region size or the correlated flux is also promising.
Despite the correlated flux at $230$~GHz being measured at the SMT-JCMT
$3.5G\lambda$ baseline, the statistics of this measurement need
to be improved towards being comparable with the statistics of total flux.
The correlated flux observations are currently being accumulated \citep{fish10}. The
correlated flux at this baseline is exponentially sensitive to the
physical flow size. As a caveat, the conclusion on image sizes may depend on the behavior of matter in the ill-defined
polar regions. Our models do not exhibit significant emission from
high latitudes at $230$~GHz (see Figure~\ref{fig_images}) or anywhere above $88$~GHz.

Future work should incorporate rigorous statistical analysis,
and such analysis should include temporal information from the observations.
The time variability properties can be found from the simulations and compared to the
observed ones. In particular, ``jet lags'' \citep{yusef_jet,maitra}
and quasi-periodic oscillations (QPOs)
\citep{genzel2003,eckart,miyoshi} should be investigated using the
simulations \citep{dolence2012}.
Also, future 3D GRMHD simulations will model more radially
extended flows, account for ADAF/ADIOS type scale-heights of
$|h/r|\sim 1$, capture outflows, and account for the effects of
accumulated magnetic flux near the black hole \citep{mtb12}.
Lastly, for the radiative transfer, adding Comptonization would be one way to test the
quiescent X-ray luminosity $L\approx4\cdot10^{32}{\rm erg~s}^{-1}$
within $2-10$~keV \citep{shcher_cond}.

\section{Acknowledgements}\label{s_ack}

The authors are grateful to Lei Huang for checking various emissivity
prescriptions, to Ramesh Narayan for extensive discussions and
comments, to Avi Loeb, Avery Broderick, James Moran, Alexander
Tchekhovskoy, Cole Miller, Julian Krolik, Steven Cranmer for
insightful comments and Jim Stone for encouragement with
self-consistent radiative transfer. We thank the anonymous referees for
their extensive feedback, which helped to improve the manuscript. The
numerical simulations and the radiative transfer calculations in this
paper were partially run on the Odyssey cluster supported by the FAS
Sciences Division Research Computing Group at Harvard, Deepthought
cluster at the University of Maryland, and were partially supported by
NSF through TeraGrid resources provided by NCSA (Abe), LONI
(QueenBee), and NICS (Kraken) under grant numbers TG-AST080025N and
TG-AST080026N. The paper is partially supported by NASA grants
NNX08AX04H (RVS\&Ramesh Narayan), NNX08AH32G (Ramesh Narayan), NASA
Hubble Fellowship grant HST-HF-51298.01 (RVS), NSF Graduate Research
Fellowship (RFP), and NASA Chandra Fellowship PF7-80048 (JCM).

\appendix
\section{RADIATIVE TRANSFER CONVERGENCE}\label{s_tests}

We have devised a novel code for GR polarized radiative transfer. As
with any new code, we need to conduct a set of convergence tests to
ensure it works accurately. First, we need to come up with metrics for
assessing accuracy. In the present paper we model fluxes at $7$
frequencies between $88$~GHz and $857$~GHz, LP fractions at $3$
frequencies and CP fractions at $2$ frequencies and define $\chi^2$ as
to characterize the goodness of fit. We employ a similar quantity
$\chi^2_H/\dof$ to characterize the accuracy of radiative transfer. We define
\begin{equation}
\chi^2_H/\dof=\frac{1}{9}\sum^{12}_{i=1}\frac{(Q_{i,1}-Q_{i,2})^2}{\sigma(Q)^2},
\end{equation}
where $Q_{i,1}$ are simulated polarized fluxes for one set of
radiative transfer parameters and $Q_{i,2}$ are the fluxes
for another set. The errors $\sigma(Q)$ are the observed errors of the
mean, and the index $i$ runs through all fitted fluxes, LP, and CP
fractions. When one of the models fits the data exactly, then
$\chi^2_H/\dof$ coincides with $\chi^2/\dof$. We vary the following radiative transfer and dynamical model parameters:
\begin{itemize}
\item{number of points $P_N$ along North-South axis and along East-West axis in the picture plane,}
\item{distance from the center $P_{\rm ss}$ measured in horizon radii $r_H$, where radiative transfer starts,}
\item{dimensionless scale $P_{\rm fact}$ of the integration region in the picture plane,}
\item{number of simulated spectra $N_{\rm periods}$ for a single model to compute the mean spectrum,}
\item{time interval $\Delta t$ of simultaneous propagation of rays and evolution of numerical simulations,}
\item{extension power-law slope of density profile $P_{\rm rhopo}$,}
\item{extension slope of temperature profile $P_{\rm Upo}$,}
\item{extension slope of magnetic field profile $P_{\rm Bpo}$.}
\end{itemize}
Since fluctuations and differences in $\chi^2/\dof$ between different
models reach $1$, then values $\chi^2_H/\dof\lesssim0.1$ are
acceptable, but, in general, we strive for $\chi^2_H/\dof<0.02$. We set
constant $P_{\rm fact}$, $P_{\rm ss}$, $P_{\rm snxy}$ for all
radiative transfer computations, but we cannot check the code accuracy
for all models. We check the convergence a posteriori for the best fit model at
each spin value.

We find values of parameters by trial-and-error. The resulting set has $P_{\rm fact}=1$, $P_{\rm ss}=1.01r_H$,
$P_{\rm snxy}=111$, $N_{\rm periods}=21$, $\Delta t=120M$. The values of $P_{\rm rhopo}$ and $P_{\rm Upo}$ are fixed by extensions to large
radii of temperature and density in the inner flow.

The tests and the values of $\chi^2_H/\dof$ are summarized in Table~\ref{tab_rad}. The second column describes
the test. In particular, $P_{\rm fact}:1\rightarrow0.8$ means that we
test convergence of the integration region relative size. We change one
parameter at a time. Since the power-law slopes $P_{\rm rhopo}$ and
$P_{\rm Upo}$ can vary from model to model, we change them in such a
way that $P_{\rm rhopo}$ is increased by $0.2$ and $P_{\rm Upo}$ is
decreased by $0.1$. We also estimate the influence of magnetic field extension power-law
slope $P_{\rm Bpo}$ by making it shallower from $(r/M)^{-1.0}$ to $(r/M)^{-0.8}$.
We chose to test relatively small variations $\Delta P_{\rm rhopo}=0.2$ and $\Delta P_{\rm Upo}=0.1$,
because density and temperature at $r_{\rm out}=3\cdot10^5M$ are known to within a factor of
several \citep{baganoff,shcher_cond}, while these variations correspond to
changes by factors of $7$ and $2.5$ in density and temperature, respectively, at $r_{\rm out}$.

\begin{table*}
\caption{Values of $\chi^2_H/\dof$ for radiative transfer convergence tests and sensitivity to model parameters tests for best fit models.}\label{tab_rad}
\begin{tabular}{| p{11mm} | p{38mm} | p{20mm}| p{20mm} | p{20mm} | p{20mm} | p{22mm} |}
\tableline\tableline
 Number & Description    & spin $a_*=0$ & spin $a_*=0.5$ & spin $a_*=0.7$ & spin $a_*=0.9$ & spin $a_*=0.98$ \\\tableline
1 & $P_{\rm N}:75\rightarrow111$                        & 0.00081 & 0.00138 & 0.00101 & 0.00047 & 0.01175 \\\tableline
2 & $P_{\rm N}:111\rightarrow161$                       & 0.00017 & 0.00072 & 0.00018 & 0.00007 & 0.00084 \\\tableline
3 & $P_{\rm ss}:1.003r_H\rightarrow 1.01r_H$            & 0.00036 & 0.00059 & 0.00110 & 0.00095 & 0.00051 \\\tableline
4 & $P_{\rm ss}:1.01r_H\rightarrow 1.03r_H$             & 0.00778 & 0.00982 & 0.01616 & 0.01468 & 0.00920 \\\tableline
5 & $P_{\rm fact}:0.8\rightarrow 1.0$                   & 0.01358 & 0.07278 & 0.06905 & 0.02893 & 0.02686 \\\tableline
6 & $P_{\rm fact}:1.0\rightarrow 1.2$                   & 0.00087 & 0.05532 & 0.07173 & 0.02681 & 0.03534 \\\tableline
7 & $N_{\rm periods}$ : 11 $\rightarrow$ 21             & 0.15665 & 0.40611 & 0.12233 & 0.21397 & 0.13588 \\\tableline
8 & $N_{\rm periods}$ : 21 $\rightarrow$ 41             & 0.02474 & 0.04505 & 0.13244 & 0.02684 & 0.04834 \\\tableline
9 & interval $\Delta t$: $120M$ $\rightarrow$ $180M$    & 0.06987 & 0.06851 & 0.13549 & 0.02948 & 0.10979 \\\tableline
10 & interval $\Delta t$: $80M$ $\rightarrow$ $120M$    & 0.09095 & 0.04103 & 0.03094 & 0.03881 & 0.06246 \\\tableline
11 & interval $\Delta t$: $0M$ $\rightarrow$ $120M$     & 0.09051 & 0.35296 & 0.53045 & 0.02731 & 0.07881 \\\tableline
12 & $P_{\rm rhopo}:Q\rightarrow Q+=0.2$                & 0.04493 & 0.04057 & 0.03134 & 0.01587 & 0.05241 \\\tableline
13 & $P_{\rm Upo}:Q\rightarrow Q-=0.1$                  & 0.01200 & 0.02726 & 0.00977 & 0.01088 & 0.04174 \\\tableline
14 & $P_{\rm Bpo}:-1.0\rightarrow -0.8$                 & 0.02401 & 1.05214 & 0.15156 & 0.05486 & 0.04941 \\\tableline
\end{tabular}
\end{table*}

The results of the tests are as follows. The first $11$ tests represent
variations of radiative transfer parameters and last $3$ tests explore the variations of power-law extension slopes.
Tests $1-4$ produce small $\chi^2_H/\dof$, so that $P_{\rm N}$ can be lowered and $P_{\rm ss}$ can be increased.
The changes in the integration region scale ($P_{\rm fact}$) result in high $\chi^2_H/\dof\approx0.07$, as indicated by tests $5$ and $6$.
Low $P_{\rm fact}$ leads to systematic underproduction of total flux, whereas high $P_{\rm fact}$ mainly leads to different $\rm LP$ fractions.
Test $7$ results in high $\chi^2_H/\dof\approx0.4$, so that a small number of simulated spectra (e.g. $N_{\rm periods}=11$) cannot be justified.
Lower values $\chi^2_H/\dof\approx0.13$ attained in test $8$ indicate that $N_{\rm periods}=21$ periods might be acceptable.
With tests $9-11$, we tested variations in the time interval $\Delta_t$ of simultaneous propagation of rays and evolution of numerical simulations.
It is expected that longer intervals lead to convergence. However, switching from $\Delta t=120M$ to $\Delta t=180M$ and switching
$\Delta t=80M$ to $\Delta t=120M$ both lead to $\chi^2_H/\dof\lesssim0.1$. Since these values of $\chi^2_H/\dof$ are acceptable, we implement $\Delta t=120M$
for radiative transfer runs. As elucidated by test $11$, freezing simulations in time leads to $\chi^2_H/\dof\approx0.5$, which is too high.
Thus, conducting radiative transfer over frozen simulation snapshots is not acceptable.
Changes in extension slopes of density and temperature (tests $12$ and $13$) result in small $\chi^2_H/\dof\lesssim0.05$.
Variations of magnetic field slope (test $14$) lead to large $\chi^2_H/\dof\approx1$, which means the modifications of
${\bf b}$ extensions will change the best fits. Extensions as shallow as
$|{\bf b}|\propto (r/M)^{-0.5}$ may provide better fits to Faraday
rotation measure and should be carefully explored. Various extensions
of the fluid velocity lead to practically the same polarized
intensities and are not included in tests.

\end{document}